\documentclass[a4paper,fleqn,usenatbib]{mnras}

\usepackage[T1]{fontenc}
\usepackage{ae,aecompl}

\usepackage{graphicx}	% Including figure files
\usepackage{amsmath}	% Advanced maths commands
\usepackage{amssymb}	% Extra maths symbols
\usepackage{subfigure}
\usepackage{multirow}
\usepackage{longtable}

\newcommand{\dinamo}{{\sc dinamo}}
\newcommand{\pcc}{\,{\rm cm}^{-3}}
\newcommand{\gcc}{\,{\rm g \, cm}^{-3}}

\newcommand{\um}{\, {\rm \mu m}}
\newcommand{\nm}{\, {\rm nm}}

\newcommand{\kel}{\, {\rm K}}

\newcommand{\msun}{\, {\rm M}_\odot}

\newcommand{\nel}{n_{\rm e}}

\newcommand{\pc}{\, {\rm pc}}

\newcommand{\nion}{n_{\rm i}}
\newcommand{\tel}{T_{\rm e}}
\newcommand{\tion}{T_{\rm i}}

\newcommand{\kpc}{\, {\rm kpc}}

\newcommand{\gfiftyfour}{G$54.1$+$0.3$}

\newcommand{\kms}{\, {\rm km \, s^{-1}}}

\newcommand{\gyr}{\, {\rm Gyr}}
\newcommand{\chsq}{\chi_{\rm red.}^2}
\newcommand{\fdest}{f_{\rm dest}}

\title[Cas A dust destruction]{Dust destruction and survival in the Cassiopeia A reverse shock}

\author[Priestley et al.]{
  F. D. Priestley$^{1}$\thanks{Email: priestleyf@cardiff.ac.uk},
  M. Arias$^2$,
  M. J. Barlow$^{3}$ \&
  I. De Looze$^{3,4}$
\\
$^{1}$School of Physics and Astronomy, Cardiff University, Queen's Buildings, The Parade, Cardiff CF24 3AA, UK \\
$^{2}$Leiden Observatory, Leiden University, P.O. Box 9513, 2300 RA Leiden, The Netherlands\\
$^{3}$Department of Physics and Astronomy, University College London, Gower Street, London WC1E 6BT, UK\\
$^{4}$Sterrenkundig Observatorium, Ghent University, Krijgslaan 281 - S9, 9000 Gent, Belgium\\
}

\date{Accepted XXX. Received YYY; in original form ZZZ}

\pubyear{2021}

\begin{document}
\label{firstpage}
\pagerange{\pageref{firstpage}--\pageref{lastpage}}
\maketitle

\begin{abstract}

Core-collapse supernovae (CCSNe) produce large ($\gtrsim 0.1 \msun$) masses of dust, and are potentially the primary source of dust in the Universe, but much of this dust may be destroyed before reaching the interstellar medium. Cassiopeia A (Cas A) is the only supernova remnant where an observational measurement of the dust destruction efficiency in the reverse shock is possible at present. We determine the pre- and post-shock dust masses in Cas A using a substantially improved dust emission model. {{In our preferred models,} the unshocked ejecta contains $0.6-0.8 \msun$ of $0.1 \um$ silicate grains, while the post-shock ejecta has $0.02-0.09 \msun$ of $5-10 \nm$ grains in dense clumps, and $2 \times 10^{-3} \msun$ of $0.1 \um$ grains in the diffuse {X-ray emitting shocked ejecta}. The implied dust destruction efficiency is $74-94 \%$ in the clumps and $92-98 \%$ overall, giving Cas A a final dust yield of $0.05-0.30 \msun$. {If the unshocked ejecta grains are larger than $0.1 \um$, the dust masses are higher, the destruction efficiencies are lower, and the final yield may exceed $0.5 \msun$.} As Cas A has a dense circumstellar environment and thus a much stronger reverse shock than is typical, the average dust destruction efficiency across all CCSNe is likely to be lower, and the average dust yield higher.} This supports a mostly-stellar origin for the cosmic dust budget.

\end{abstract}


%%%%%%%%%%%%%%%%%%%%%%%%%%%%%%%%%%%%%%%%%%%%%%%%%%

%%%%%%%%%%%%%%%%% BODY OF PAPER %%%%%%%%%%%%%%%%%%

\section{Introduction}

The presence of large masses of interstellar dust in high-redshift galaxies \citep{bertoldi2003,priddey2003,watson2015}, less than a $\gyr$ after the Big Bang, requires a mechanism for rapid dust production. Core-collapse supernovae (CCSNe) have been suggested as potential sources of this dust, which would require an average dust yield of $\gtrsim 0.1 \msun$ per SN to explain the observationally derived dust masses \citep{morgan2003,dwek2007}. Far-infrared (IR) observations of supernova remnants (SNRs) have detected quantities of cold ejecta dust comparable to this value \citep{matsuura2015,delooze2017,delooze2019,chawner2019,chawner2020}, as have investigations of spectral line asymmetries \citep{bevan2017,bevan2019,bevan2020}, but the final dust yield enriching the interstellar medium (ISM) depends on how much of this newly-formed dust can survive the reverse shock. With a reverse-shock dust destruction efficiency of $\fdest \lesssim 50 \%$, \citet{delooze2020} find that CCSNe are the dominant producers of dust over the history of the Universe, whereas \citet{galliano2021}, with {a maximum SN dust yield of $\sim 0.03 \msun$ and best-fit value of $7 \times 10^{-3} \msun$} (corresponding to $\fdest \gtrsim 90\%$ for typical observed dust masses), find them to be unimportant except at very low metallicity. Both studies treat $\fdest$ as a free parameter (implicitly, in the case of \citealt{galliano2021}) when fitting chemical evolution models to observed galaxy properties; an independent constraint on $\fdest$ is necessary in order to break model degeneracies and further our understanding of cosmic dust evolution.

Simulations of dust destruction in SNR reverse shocks vary from almost complete destruction of the newly-formed dust to almost complete survival; \citet{kirchschlager2019} find reported values in the literature ranging from $\fdest \sim 1-100 \%$, with their own values falling towards the higher end of this range. The results are highly sensitive to the choice of input physics, both relating to the large-scale hydrodynamics of the SNR, and the microphysics of the dust grains. While grains with radii $a \gtrsim 0.1 \um$ were previously thought to be resistant to sputtering, and would thus mostly survive into the ISM \citep{nozawa2007}, the inclusion of shattering in grain-grain collisions makes $\fdest$ strongly (and non-linearly) dependent on the initial size distribution and grain composition \citep{kirchschlager2019}. Additional complications such as departures from spherical symmetry in the ejecta \citep{kirchschlager2019,slavin2020} and a potentially non-uniform surrounding medium \citep{martinezgonzales2019} have only recently begun to be addressed, with no clear consensus on the value (or range of values) $\fdest$ should take.

Cassiopeia A (Cas A) is one of few SNRs with a developed reverse shock, and is certainly the best studied of those. \citet{barlow2010} found that the far-IR spectral energy distribution (SED) required the presence of $\gtrsim 0.1 \msun$ of dust in the SN ejecta. Using a more complex, {spatially resolved} dust model, accounting for unrelated foreground emission from the ISM and multiple SNR dust temperature components, \citet{delooze2017} found $\sim 0.6 \msun$ of cold ($\sim 30 \kel$) silicate grains located within the radius of the reverse shock, in addition to a smaller mass ($\sim 0.01 \msun$) of warmer dust. In \citet{priestley2019}, we used models of dust heating in the various ejecta components to fit the IR SED and determine the pre- and post-shock dust masses in Cas A. Combined with estimates of the corresponding gas masses, we found dust-to-gas (DTG) mass ratios of $0.2$ and $0.1$ in the pre- and post-shock ejecta clumps respectively, providing an observational estimate of $\fdest \sim 50 \%$ for this object. However, several of the assumptions underlying this result have since been cast into some doubt, particularly regarding the grain size distribution \citep{priestley2020} and the gas mass in the unshocked ejecta \citep{laming2020}. In this paper, we revisit the \citet{priestley2019} model of the Cas A dust emission, making numerous improvements to both the physics and the fitting procedure, in order to derive an updated observational estimate of $\fdest$ for the only object for which this is presently feasible.

\section{Method}
\label{sec:method}

\begin{figure*}
  \centering
  \includegraphics[width=2\columnwidth]{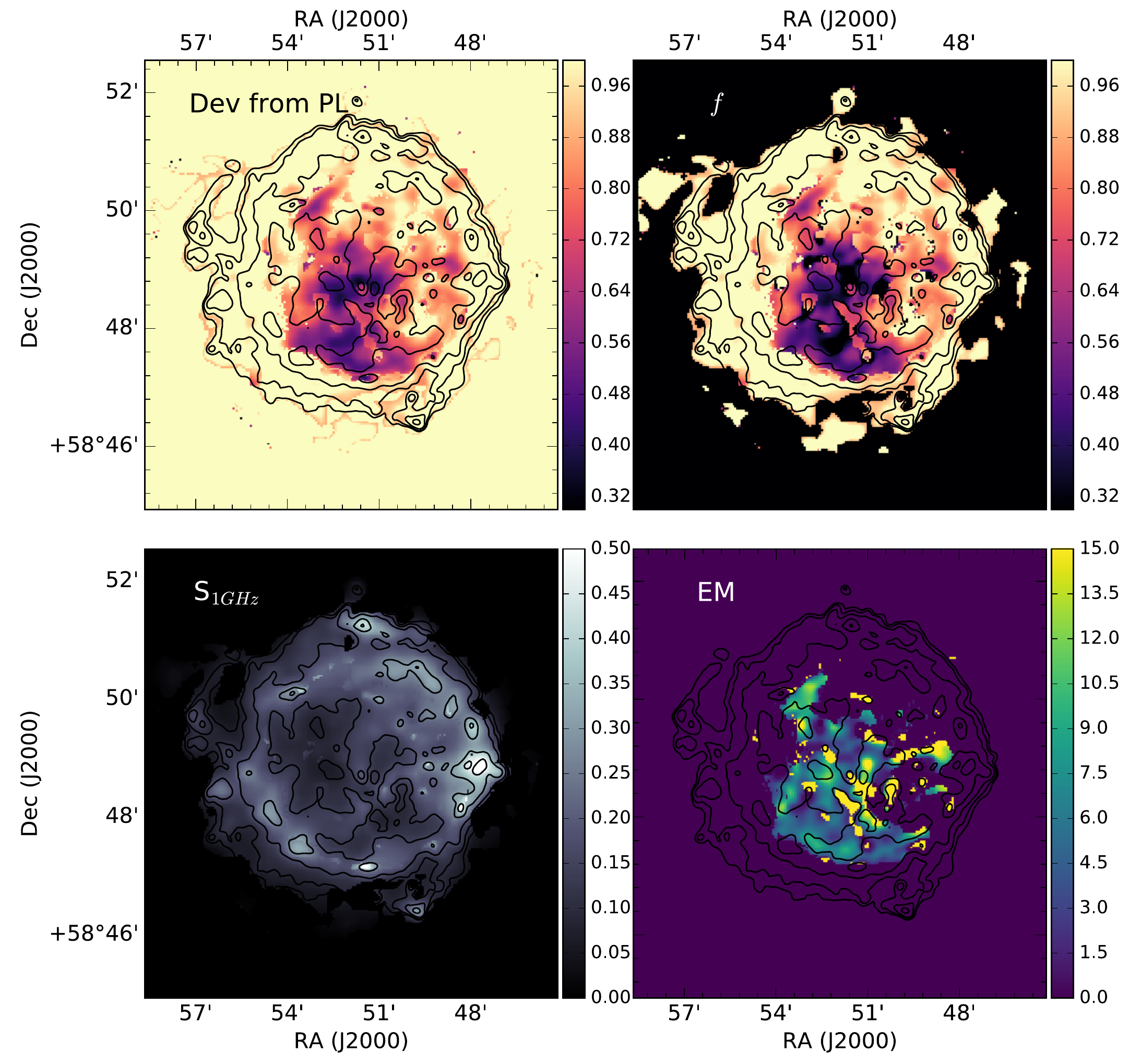}
  \caption{Results of fitting the Cas A narrow-band images from  \citet{degasperin2020} to equations 5 to 7 in \citet{arias2018}. $S_{1\mathrm{GHz}}$ is the best-fit flux density at 1~GHz in Jy, and Dev from PL is the best-fit deviation from a power-law spectrum with $S_\nu \propto S_{1\mathrm{GHz}}(\frac{\nu}{1~\mathrm{GHz}})^{-0.77}$. $EM$ is the emission measure from the absorbing material given in pc~cm$^{-6}$, and $f$ is the fraction of material lying in front of the absorbing material (and that is therefore not subject to free-free absorption).}
  \label{fig:absresults}
\end{figure*}

\begin{table*}
  \centering
  \caption{Adopted gas masses, electron and ionic densities and temperatures, and dominant ionic species, for the three components in Cas A. References: (1) This work (2) \citet{smith2009} (3) \citet{raymond2018} (4) \citet{priestley2019} (5) \citet{docenko2010} (6) \citet{willingale2003}}
  \begin{tabular}{cccccccc}
    \hline
    Component & $M_{\rm gas} / \msun$ & $\nion / \pcc$ & $\nel / \pcc$ & $\tion / \kel$ & $\tel / \kel$ & Ion & Ref. \\ 
    \hline
    Preshock & $0.53$ & $100$ & $100$ & $100$ & $100$ & O & (1),(2),(3) \\
    Clumped & $0.59$ & $480$ & $480$ & $10^4$ & $10^4$ & O & (4),(5) \\
    Diffuse & $1.68$ & $7.8$ & $61$ & $7.05 \times 10^8$ & $5.22 \times 10^6$ & O & (6) \\
    \hline
  \end{tabular}
  \label{tab:gasprop}
\end{table*}

\begin{table}
  \centering
  \caption{Cas A SNR dust IR fluxes for the $G=0.6$ ISM dust model of \citet{delooze2017}, {and those from an} annulus with inner/outer radii of $140''/165''$. Effective wavelengths for each filter are given in $\um$.}
  \begin{tabular}{cccc}
    \hline
    Filter & $F_{\nu}$ / Jy & $F_{\nu} (> 140'')$ / Jy \\
    \hline
    IRAC 8 & $0.2 \pm 0.1$ & $<0.4$ \\
    WISE 12 & $3.4 \pm 0.3$ & $<0.3$ \\
    IRS 17 & $63.3 \pm 6.0$ & $10.6 \pm 2.3$ \\
    WISE 22 & $202.0 \pm 19.3$ & $27.7 \pm 2.6$ \\
    MIPS 24 & $154.4 \pm 15.0$ & $22.0 \pm 1.7$ \\
    IRS 32 & $168.5 \pm 17.3$ & $25.3 \pm 5.0$ \\
    PACS 70 & $149.5 \pm 20.1$ & $19.3 \pm 3.0$ \\
    PACS 100 & $125.8 \pm 19.9$ & $14.6 \pm 5.0$ \\
    PACS 160 & $69.9 \pm 12.0$ & $1.1 \pm 6.9$ \\
    SPIRE 250 & $27.3 \pm 4.8$ & $<3.8$ \\
    SPIRE 350 & $10.9 \pm 1.9$ & $<2.5$ \\
    SPIRE 500 & $2.6 \pm 0.5$ & $<1.5$ \\
    SCUBA 850 & $0.4 \pm 0.1$ & - \\
    \hline
  \end{tabular}
  \label{tab:irflux}
\end{table}

\subsection{Physical conditions in Cas A}

In \citet{priestley2019}, we divided the Cas A SNR into four physical components: the unshocked ejecta (`preshock'), the clumped ejecta that has passed through the reverse shock (`clumped'), the lower-density, X-ray emitting shocked ejecta (`diffuse'), and the circumstellar material (CSM) heated by the forward shock {(`blastwave'). The diffuse and blastwave components produced very similar dust SEDs peaking in the mid-IR, which the available data in this region cannot reliably distinguish. In the updated analysis we assume the CSM contains no dust, and therefore ignore the blastwave component. Dust is far more likely to be present in the highly metal-enriched ejecta than in the mostly-hydrogen CSM, and preexisting CSM grains may well have been sublimated by radiation from the SN itself. If we make the opposite assumption (i.e. include a blastwave component at the expense of the diffuse one), our main results are virtually unchanged, due to the low dust mass required to fit the mid-IR SED in either case.} We use the gas properties (density, temperature and composition) from \citet{priestley2019}, and the same gas masses for all but the preshock component, which we discuss below; these are listed in Table \ref{tab:gasprop}. We assume a distance to the SNR of $3.4 \kpc$ \citep{reed1995}.

The preshock gas mass of $3 \msun$ used in \citet{priestley2019} was taken from the radio self-absorption study by \citet{arias2018}. It has since come to our attention that this value was dominated by a few degenerate pixels that contributed disproportionately to the mass estimate. We have redone the analysis using the complete LOFAR Cas A LBA dataset presented in \citet{degasperin2020}, with twice the observing time and five times the bandwidth of the original study, and using a more stringent mask of degenerate pixels. As intermediate products for the broad-band imaging, \citet{degasperin2020} produced 61 narrow-band images of Cas A in the frequency range of $30-77$~MHz. We followed the same method presented in sec. 3 of \citet{arias2018} and fit for deviation from power-law behaviour (assuming $\alpha=0.77$ for Cas A), coming to the results in Figure \ref{fig:absresults}.

Going from the best-fit value of the {emission measure, $4.2 \pc \, {\rm cm^{-6}}$,} to an estimate of the unshocked mass inside of Cas A's reverse shock depends on the temperature, ionisation, and geometry of the absorbing material. In \citet{arias2018} it was assumed that the absorbing material has a temperature of 100~K, an average ionisation state $Z=3$, and that the unshocked ejecta are mostly composed of oxygen, with a mass number $A$ of 16. Using these values, we arrive at the following mass estimate:
\begin{equation}
\begin{split}
M = \, & 0.53 \pm 0.10 \msun \left(\frac{A}{16}\right) \left(\frac{l}{0.16~\mathrm{pc}}\right)^{-3/2} \left(\frac{Z}{3}\right)^{-3/2}\\
&\times \left(\frac{T}{100~\mathrm{K}}\right)^{3/4} \sqrt{\frac{g_\mathrm{ff}(T=100~\mathrm{K}, Z=3)}{g_\mathrm{ff}(T, Z)}};
\end{split}
\label{eq:mass}
\end{equation}
here $g_\mathrm{ff}$ is a Gaunt factor given in eq. 2 of \citet{arias2018}. {The corresponding electron density is $\nel = 5.12 \left( \frac{l}{0.16 \pc} \right)^{-1/2} \pcc$.} The choice of temperature, ionisation state, and geometry follow the assumptions made in \citet{delaney2014}. The $T=100$~K assumption was later confirmed by \citet{raymond2018}. Moreover, we assume that the unshocked ejecta is composed mostly of {three-times ionised oxygen ([O IV], as observed by \citet{isensee2010} among others), whereas we know that there are low ionisation, heavier species, such as [Si~II] and [S~III] \citep{smith2009, isensee2010, milisavljevic2015}, and possibly also Fe.} There is also neutral material inside the reverse shock \citep{koo2018} that does not contribute to the ionisation, and therefore is not included in this estimate. Perhaps more importantly, we assume that the unshocked ejecta are confined to a disk of thickness $l=0.16$~pc. \citet{milisavljevic2015} found in fact that the internal ejecta are bubble-like, formed by cavities and ejecta walls that separate them. Finally, as discussed in \citet{arias2018}, we do not know how clumped the ejecta are. These factors can significantly increase (or, in the case of high clumping, decrease) the mass estimate as derived from radio free-free absorption. As our value agrees with the recent independent estimate of $0.49^{+ 0.47}_{-0.24} \msun$ from \citet{laming2020}, we use it throughout the rest of this paper, but note that it remains a significant source of systematic uncertainty in the estimate of $\fdest$.

In \citet{priestley2019}, the assumed shape of the radiation field responsible for dust heating was taken from \citet{wang2016}. This fit to the radio and X-ray observations included no data at the optical/ultraviolet (UV) wavelengths which typically dominate dust heating. To better constrain the radiation field, we instead use the SED of the $1000 \kms$ V\_n10\_b0.001 model from the \citet{allen2008} collection of radiative shock models, which produces post-shock properties best matching those of the blastwave from \citet{willingale2003}. We scale this to match the total X-ray luminosity of $4 \times 10^{37} \, {\rm erg \, s^{-1}}$ \citep{laming2020}, approximating the emitting surface area as a shell of radius $1.7 \pc$ \citep{reed1995}. By symmetry, the flux at any point inside a spherical shell of emitting material is identical; we can thus use the radiation field at the centre, and ignore the fact that the dust grains will in fact be located at a range of radii.

\subsection{Dust SED models}

We model the emission from dust grains using \dinamo{} \citep{priestley2019}, a dust heating and emission code accounting for stochastic heating effects on small grains, under the conditions of the three SNR components described above. Rather than assuming a \citet{mathis1977} ISM grain size distribution for each component, as in \citet{priestley2019}, we calculate single-grain SEDs for a range of grain sizes. There is strong observational evidence that dust grains formed in CCSNe are large ($a \gtrsim 0.1 \um$; \citealt{gall2014,wesson2015,bevan2016,priestley2020}), and that the size distributions are much more top-heavy that those found by \citet{mathis1977} \citep{wesson2015,owen2015,priestley2020}, but processing by the reverse shock can significantly alter the original distribution \citep{kirchschlager2019}. The available IR data are insufficient to constrain the size distribution in each component, so we assume a single grain size for each.

\citet{delooze2017} and \citet{priestley2019} both find that the dust must be mostly comprised of silicates. While the exact composition is unknown, the differences in dust masses for different silicate optical properties are relatively minor (with the exception of the Mg$_{0.7}$SiO$_{2.7}$ grains from \citet{jaeger2003}, which require unrealistically large masses), so we use the values for MgSiO$_3$ grains from \citet{dorschner1995}, extended to UV and X-ray wavelengths with data from \citet{laor1993}. We assume a bulk grain density of $2.5 \gcc$ for consistency with previous work.

We fit the $G=0.6$ Cas A SN dust SED from \citet{delooze2017}, given in Table \ref{tab:irflux}, with the number (and, equivalently, mass) of grains in each ejecta component as our three free parameters, treating the $8$ and $12 \um$ fluxes as upper limits due to the highly uncertain contribution from ISM dust at these wavelengths. Rather than performing a grid search for the minimum $\chsq$ as in \citet{priestley2019}, we fit the SED using {\it emcee} \citep{foreman2013}, a Monte Carlo Markov chain (MCMC) code, which allows us to more accurately determine the best-fit dust masses while also accounting for any model degeneracies and the observational uncertainties. We use $500$ walkers {with $5000$ steps per walker and $500$ burn-in steps}, which we have confirmed is sufficient for the models to converge. {We additionally consider an SED extracted {in the same way as the \citet{delooze2017} dust SED, but} from an annulus with inner and outer radii of $140''$ and $165''$, representing a `post-shock' SED with minimal contribution from the central, unshocked ejecta dust, {also given in Table \ref{tab:irflux}}.}

\section{Results}

\begin{figure}
  \centering
  \includegraphics[width=\columnwidth]{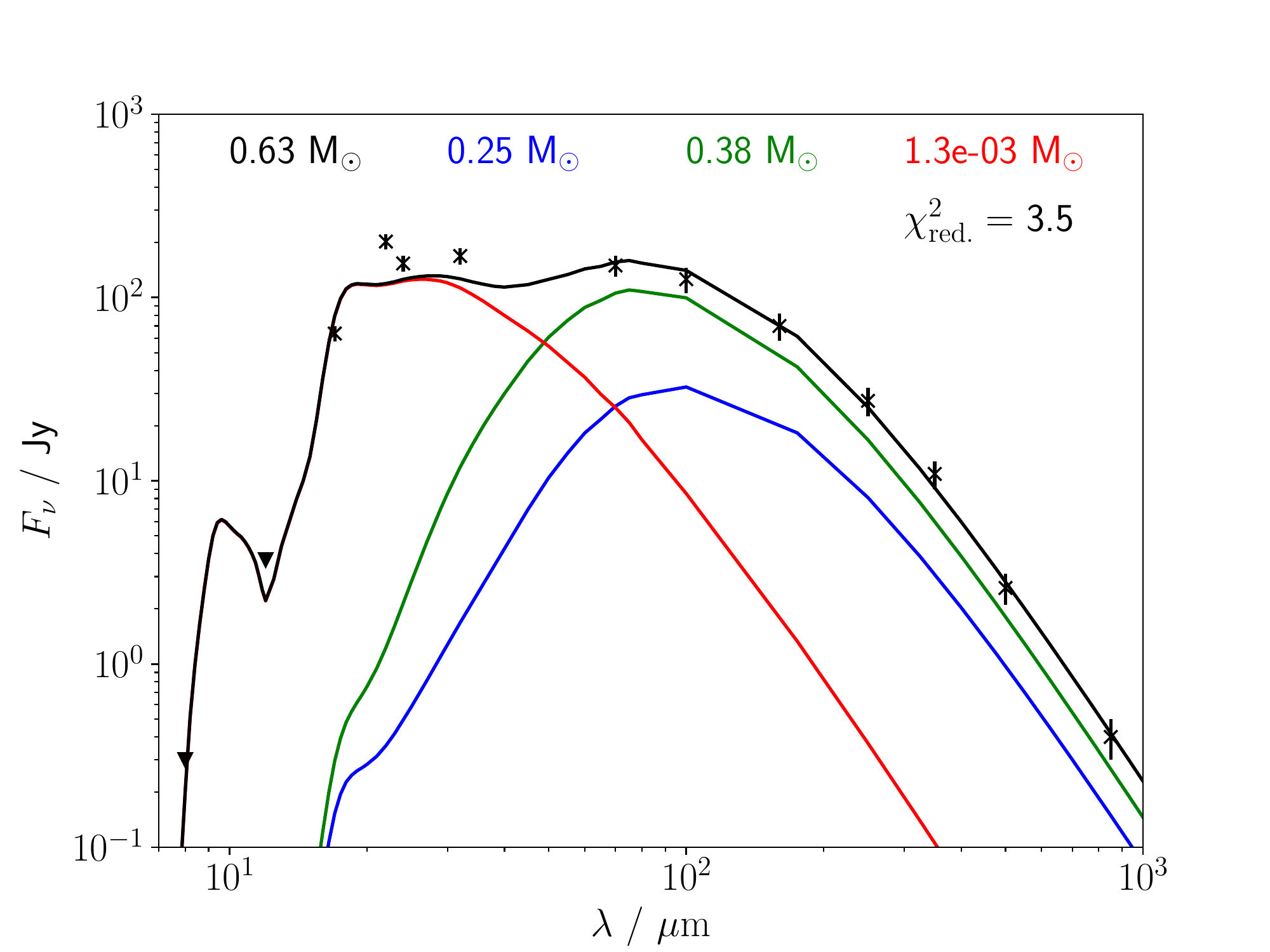}
  \caption{Cas A dust SED (black crosses) and best-fit model SEDs for preshock (blue), clumped (green), and diffuse (red) components using a \citet{mathis1977} grain size distribution as in \citet{priestley2019}, with the total model SED shown in black.}
  \label{fig:mrn}
\end{figure}

\begin{figure*}
  \centering
  \subfigure{\includegraphics[width=\columnwidth]{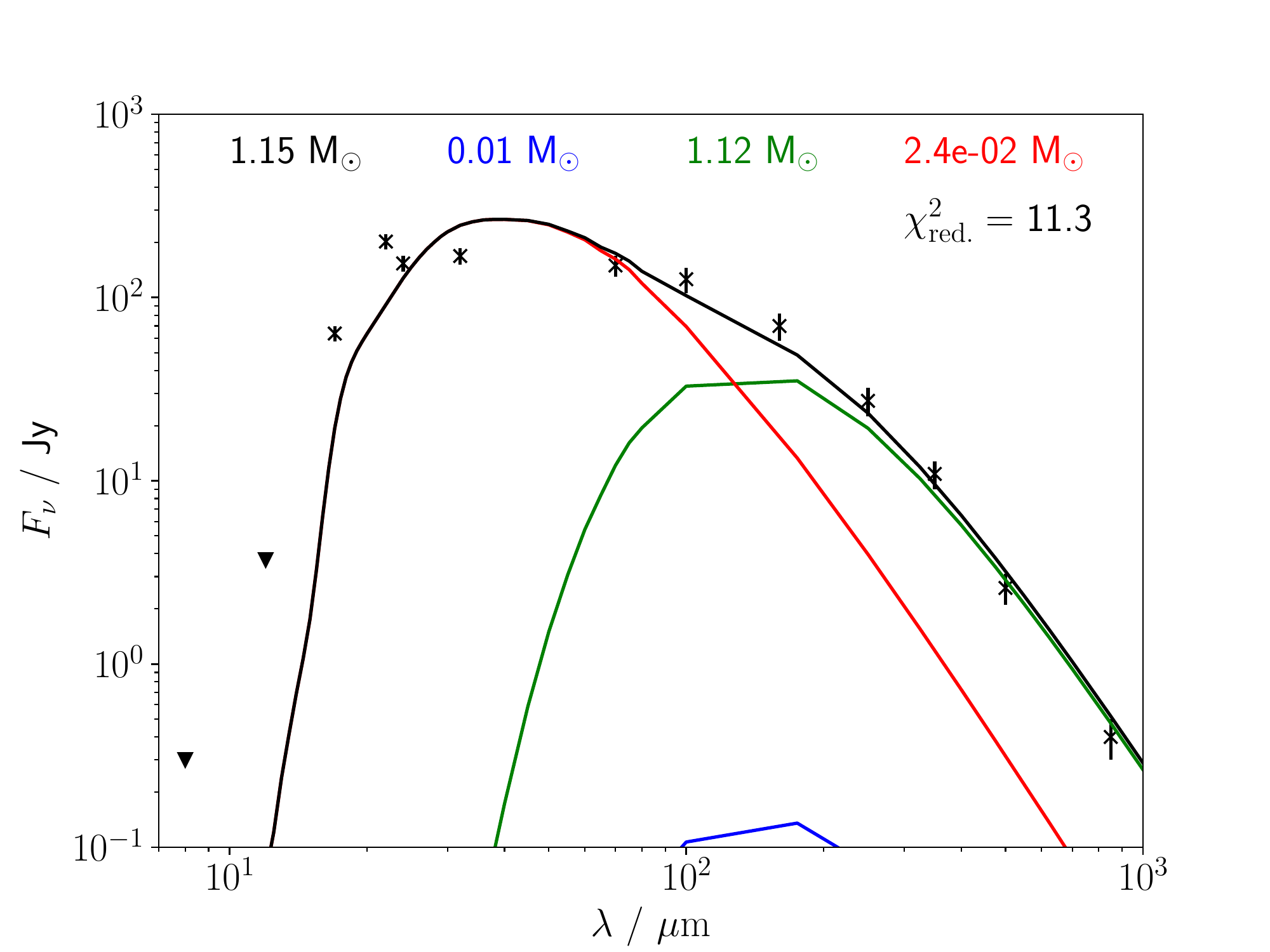}}\quad
  \subfigure{\includegraphics[width=\columnwidth]{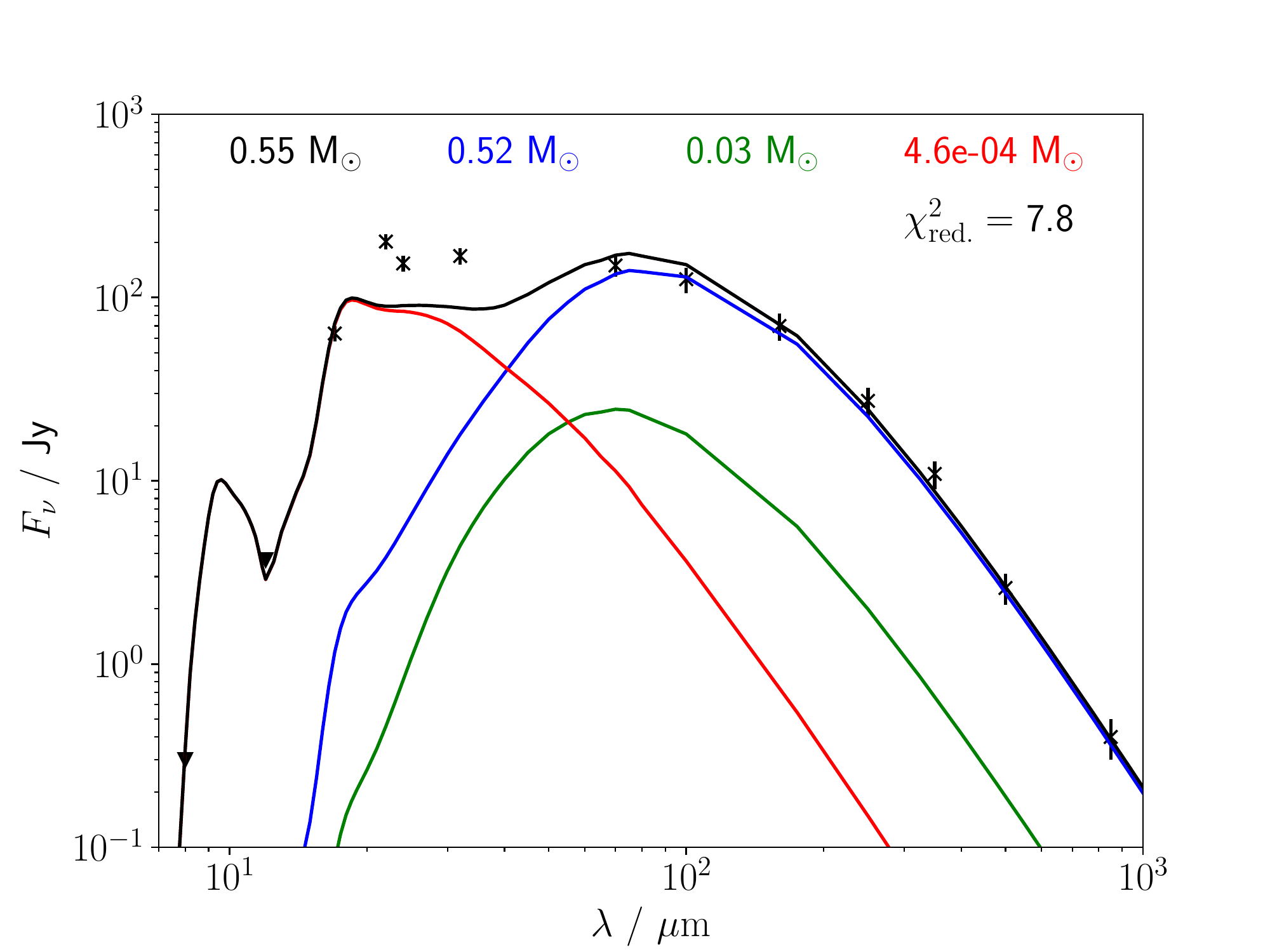}}
  \caption{Cas A dust SED (black crosses) and best-fit model SEDs for preshock (blue), clumped (green) and diffuse (red) grains of radius $1 \um$ (left, model A) or $10 \nm$ (right, model B), with the total model SEDs shown in black.}
  \label{fig:diffsizes}
\end{figure*}

\begin{figure}
  \centering
  \includegraphics[width=\columnwidth]{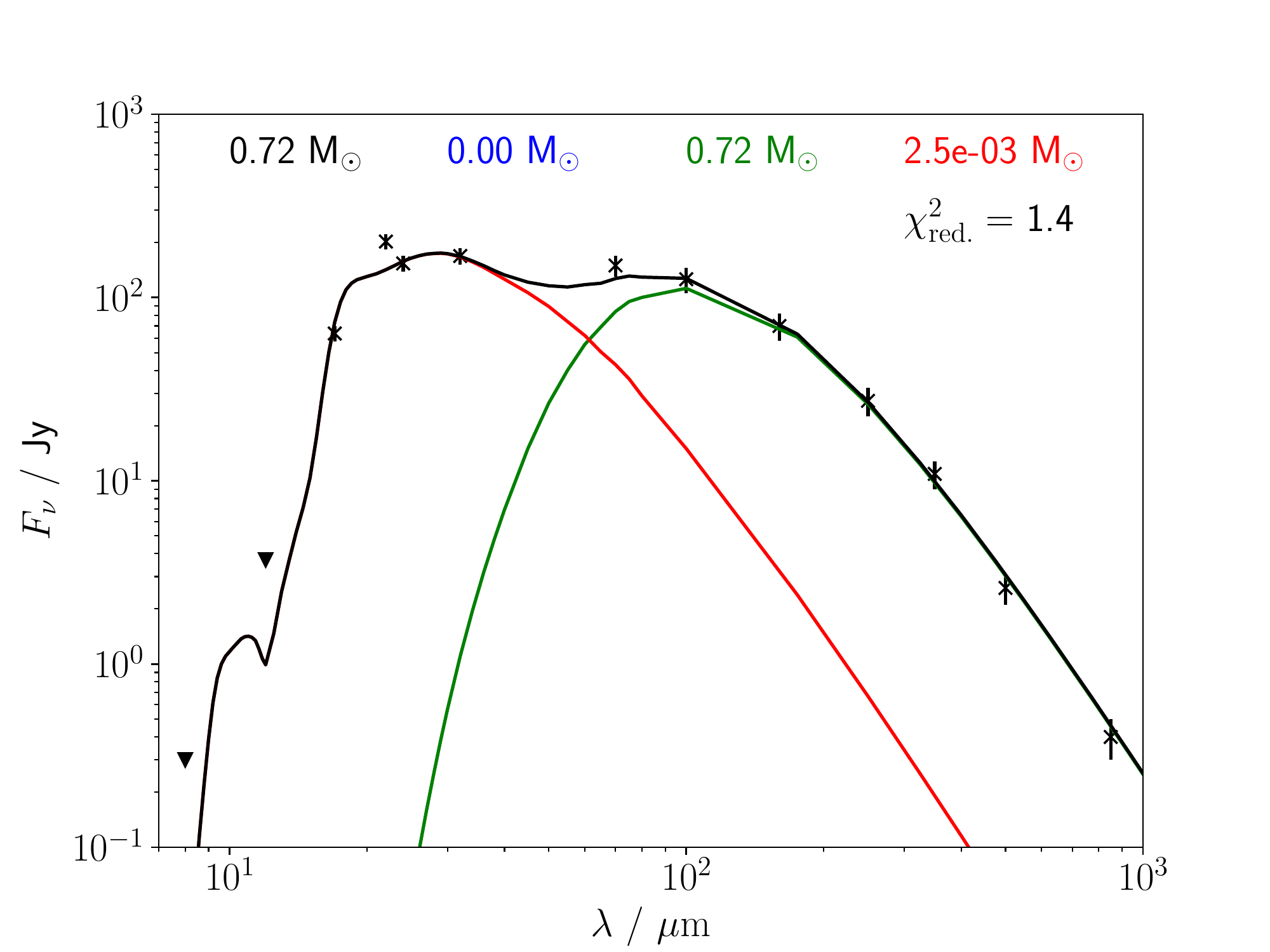}
  \caption{Cas A dust SED (black crosses) and the best-fit model SED for preshock (blue), clumped (green) and diffuse (red) grains of radius $0.1 \um$ (model C), with the total model SEDs shown in black.}
  \label{fig:all01}
\end{figure}

\begin{figure}
  \centering
  \includegraphics[width=\columnwidth]{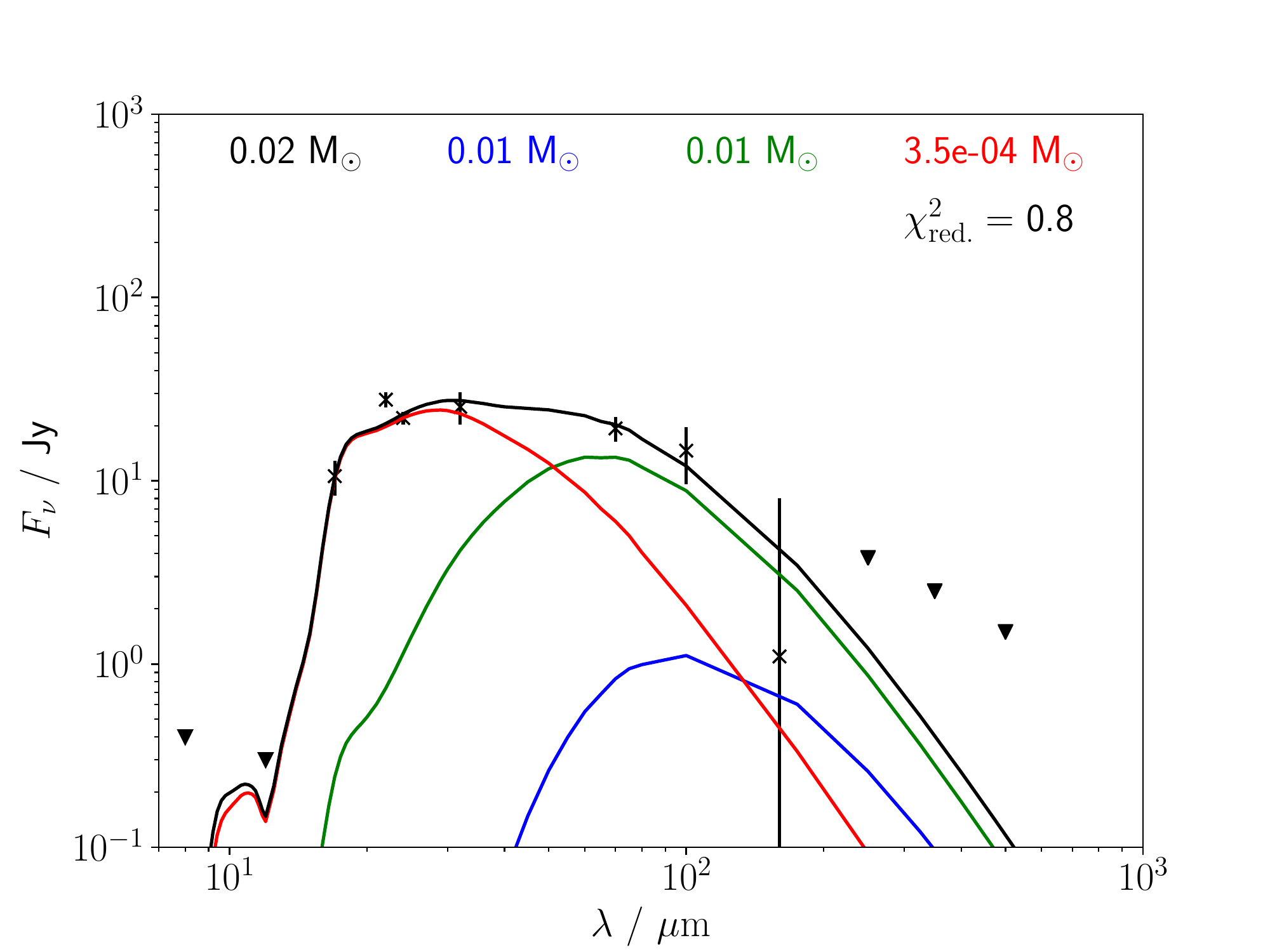}
  \caption{Cas A post-shock dust SED (black crosses) and best-fit model SEDs for {clumped} $0.1 \um$ (blue), clumped $5 \nm$ (green), and diffuse $0.1 \um$ (red) grains, with the total model SED shown in black.}
  \label{fig:postshock}
\end{figure}

\begin{figure*}
  \centering
  \subfigure{\includegraphics[width=\columnwidth]{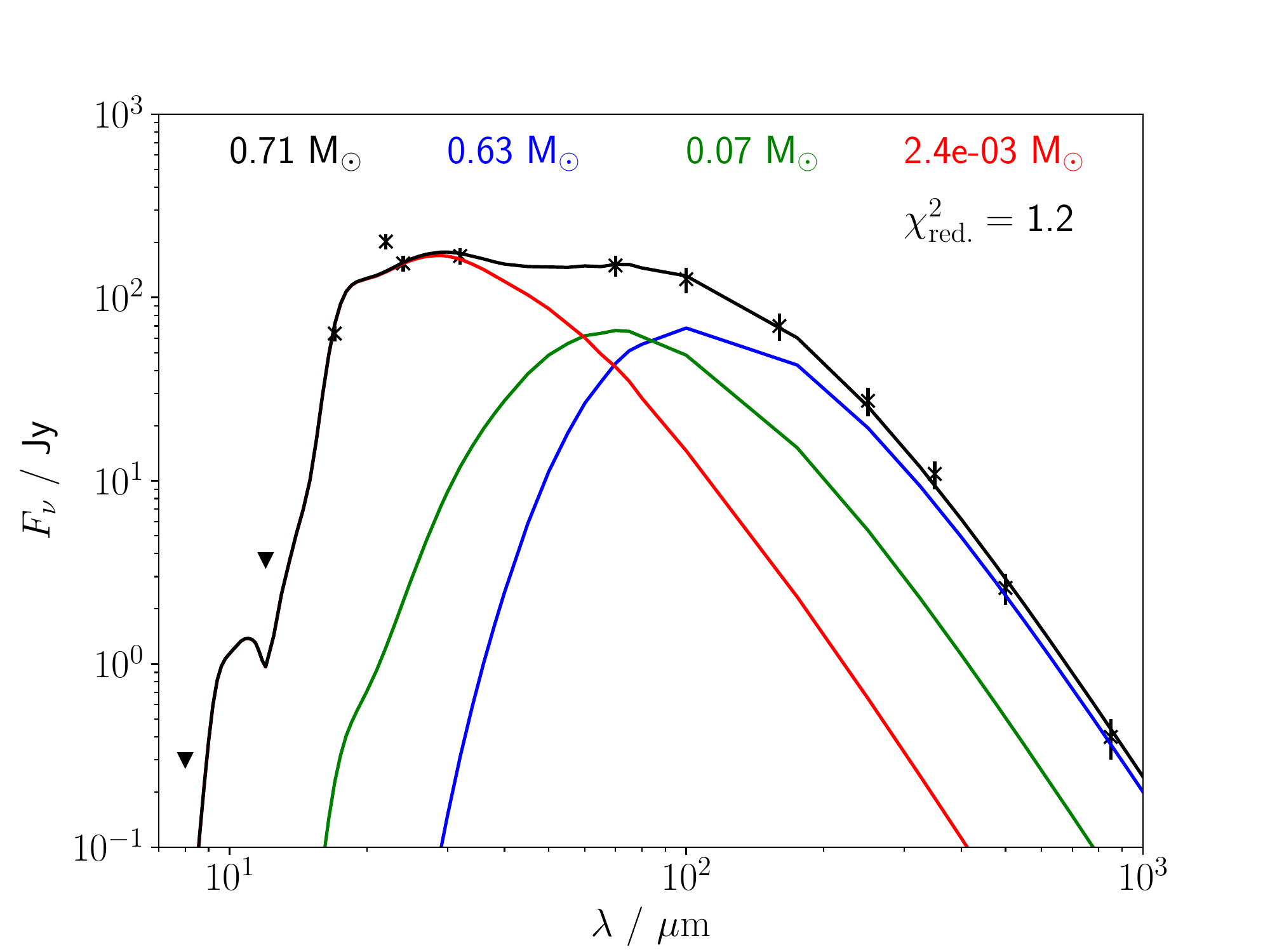}}\quad
  \subfigure{\includegraphics[width=\columnwidth]{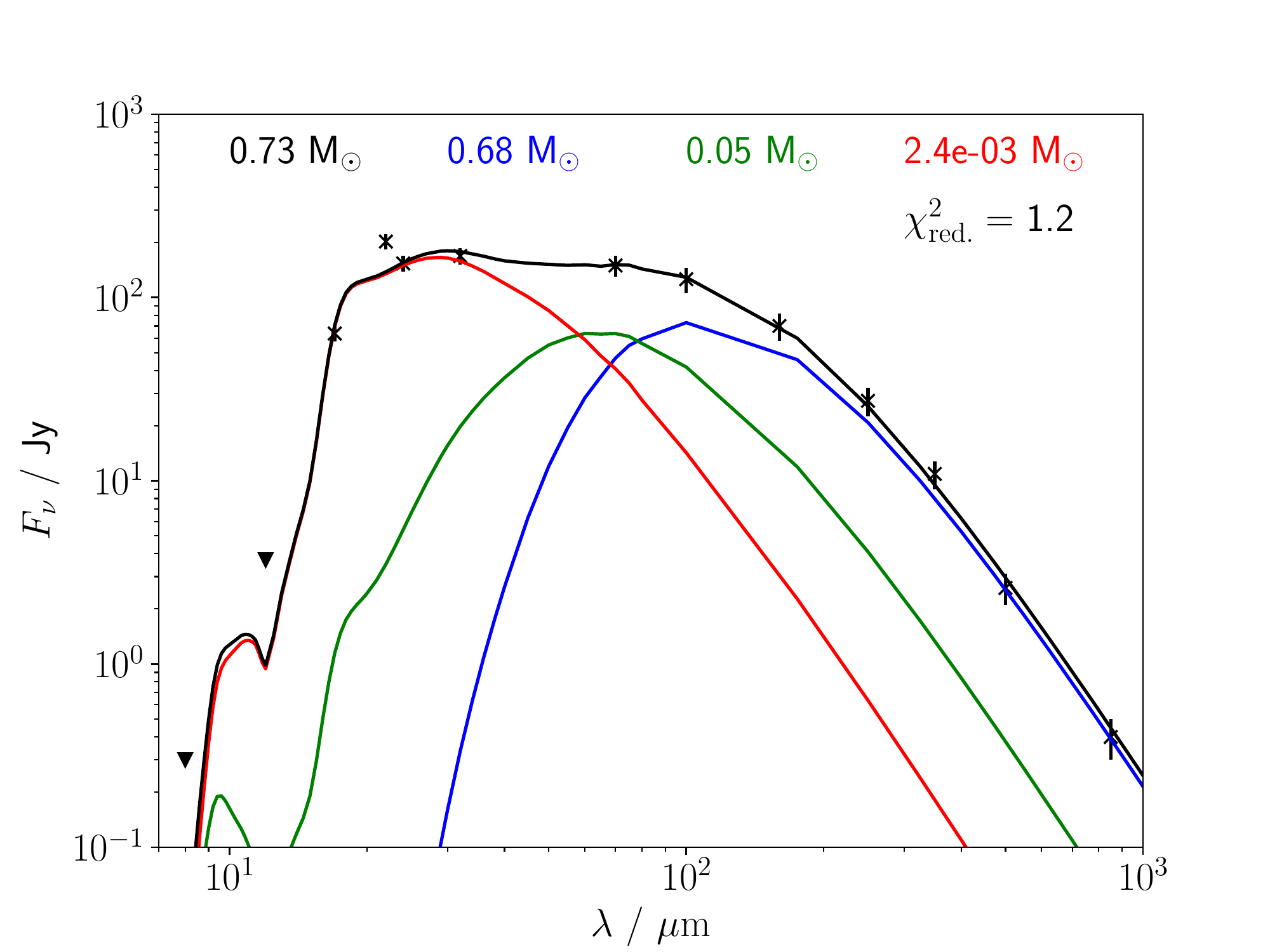}}
  \caption{Cas A dust SED (black crosses) and best-fit model SEDs for preshock $0.1 \um$ (blue) and diffuse $0.1 \um$ (red) grains, with a clumped grain size of $10 \nm$ (green, left; model D) or $5 \nm$ (green, right; model E) with the total model SED shown in black.}
  \label{fig:clumpsmall}
\end{figure*}

\begin{figure*}
  \centering
  \subfigure{\includegraphics[width=\columnwidth]{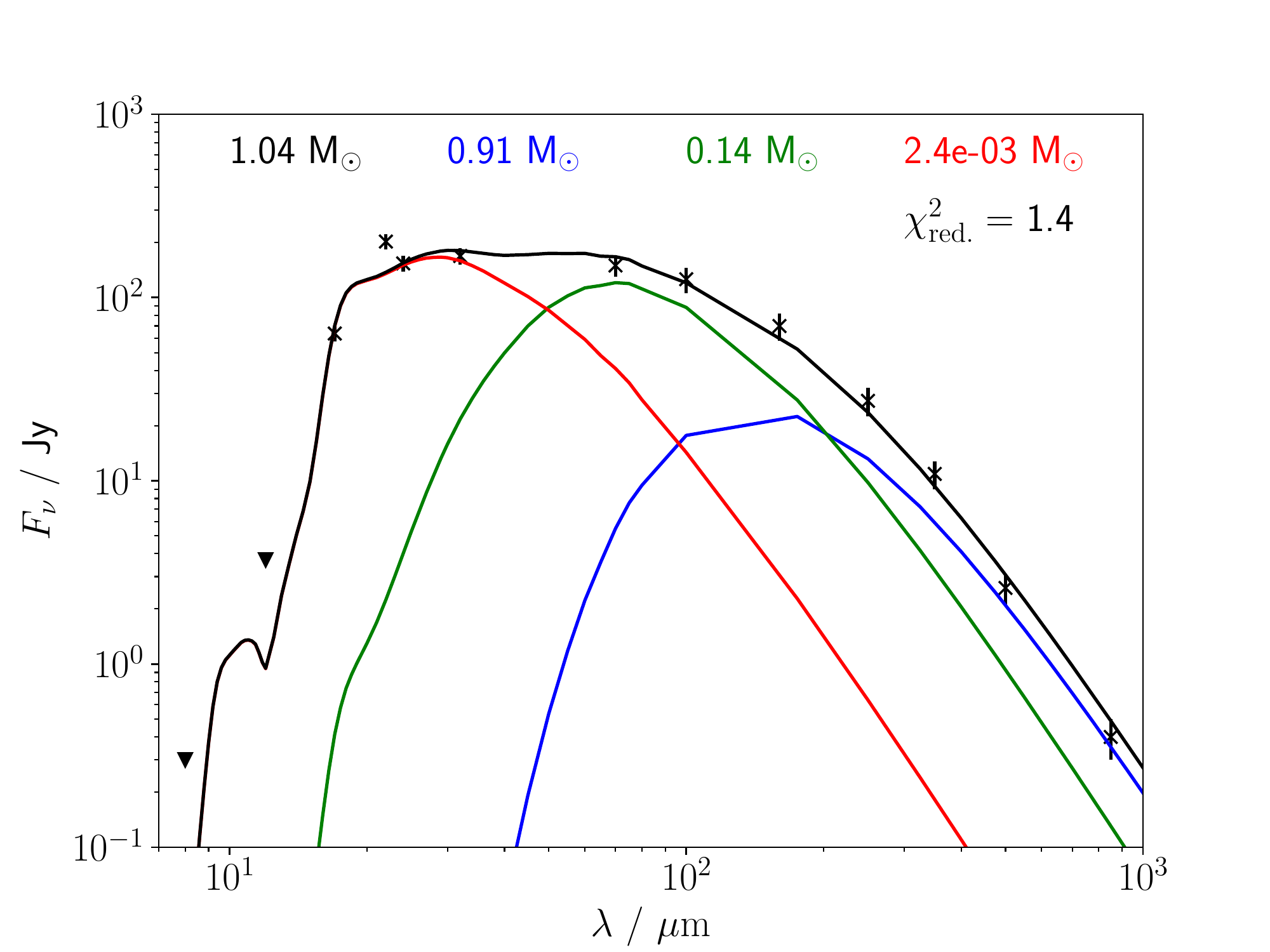}}\quad
  \subfigure{\includegraphics[width=\columnwidth]{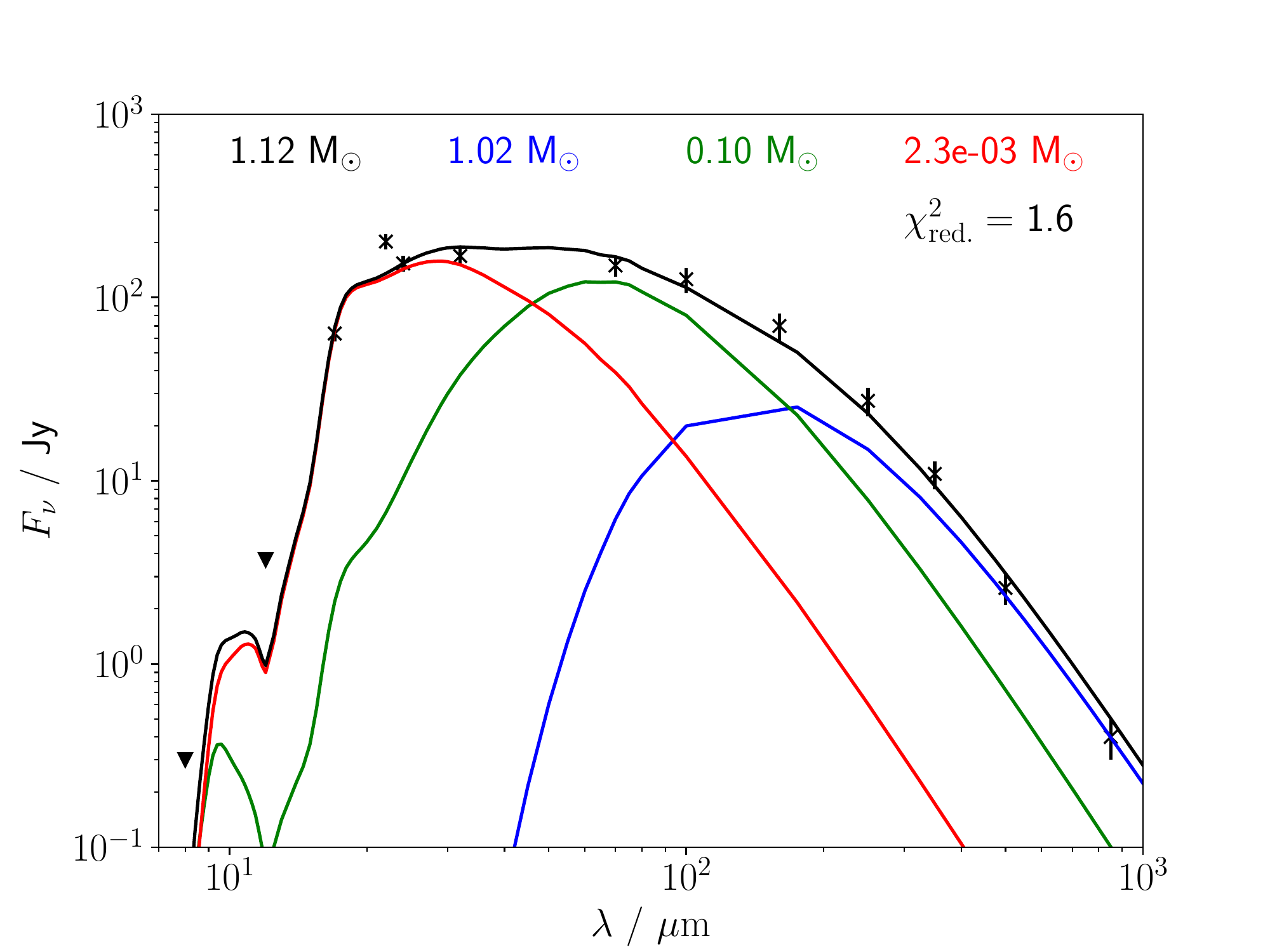}}
  \caption{Cas A dust SED (black crosses) and best-fit model SEDs for preshock $1 \um$ (blue) and diffuse $0.1 \um$ (red) grains, with a clumped grain size of $10 \nm$ (green, left; model F) or $5 \nm$ (green, right; model G) with the total model SED shown in black.}
  \label{fig:preshockbig}
\end{figure*}

\begin{figure*}
  \centering
  \includegraphics[width=\columnwidth]{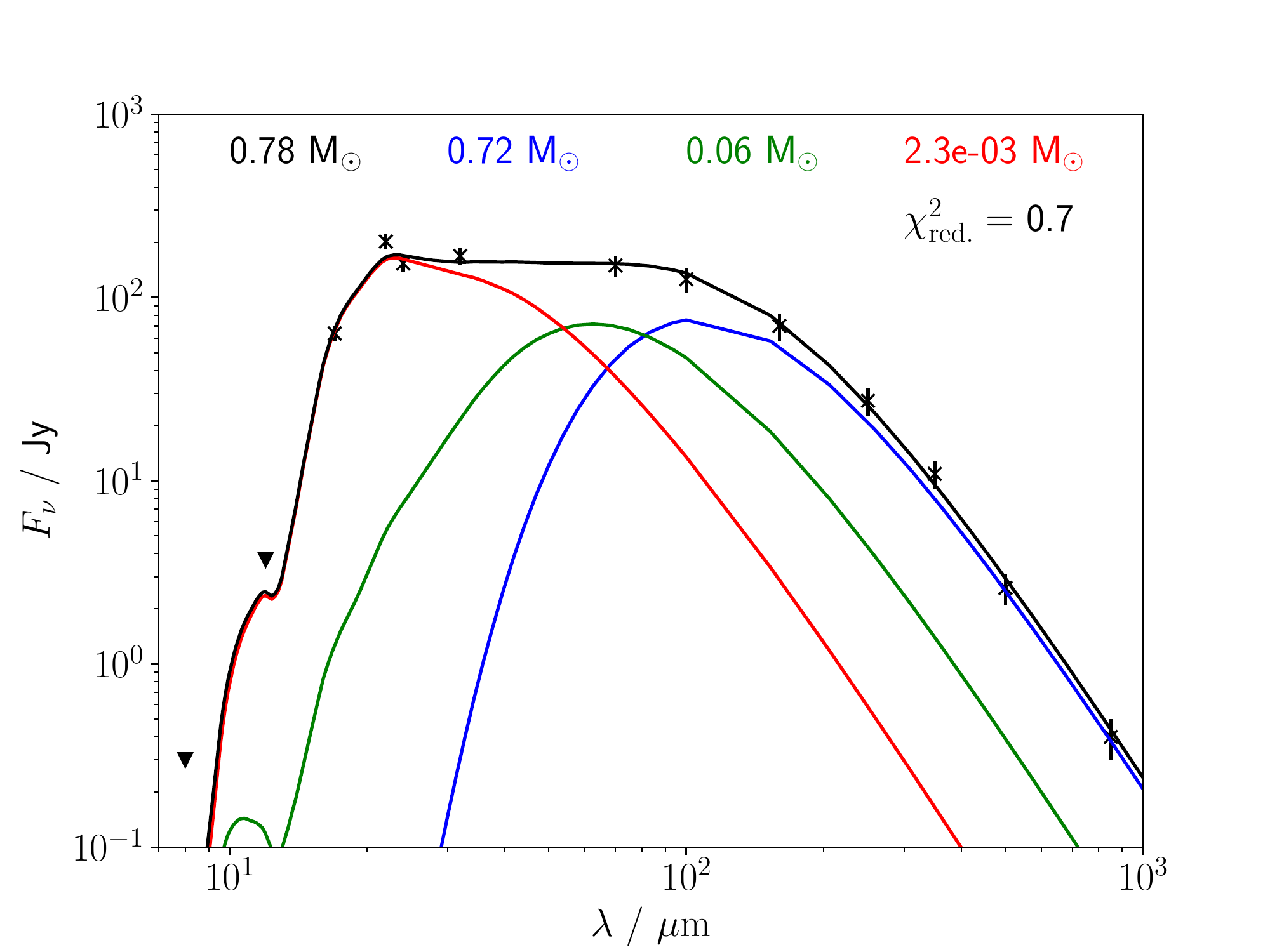}
  \includegraphics[width=\columnwidth]{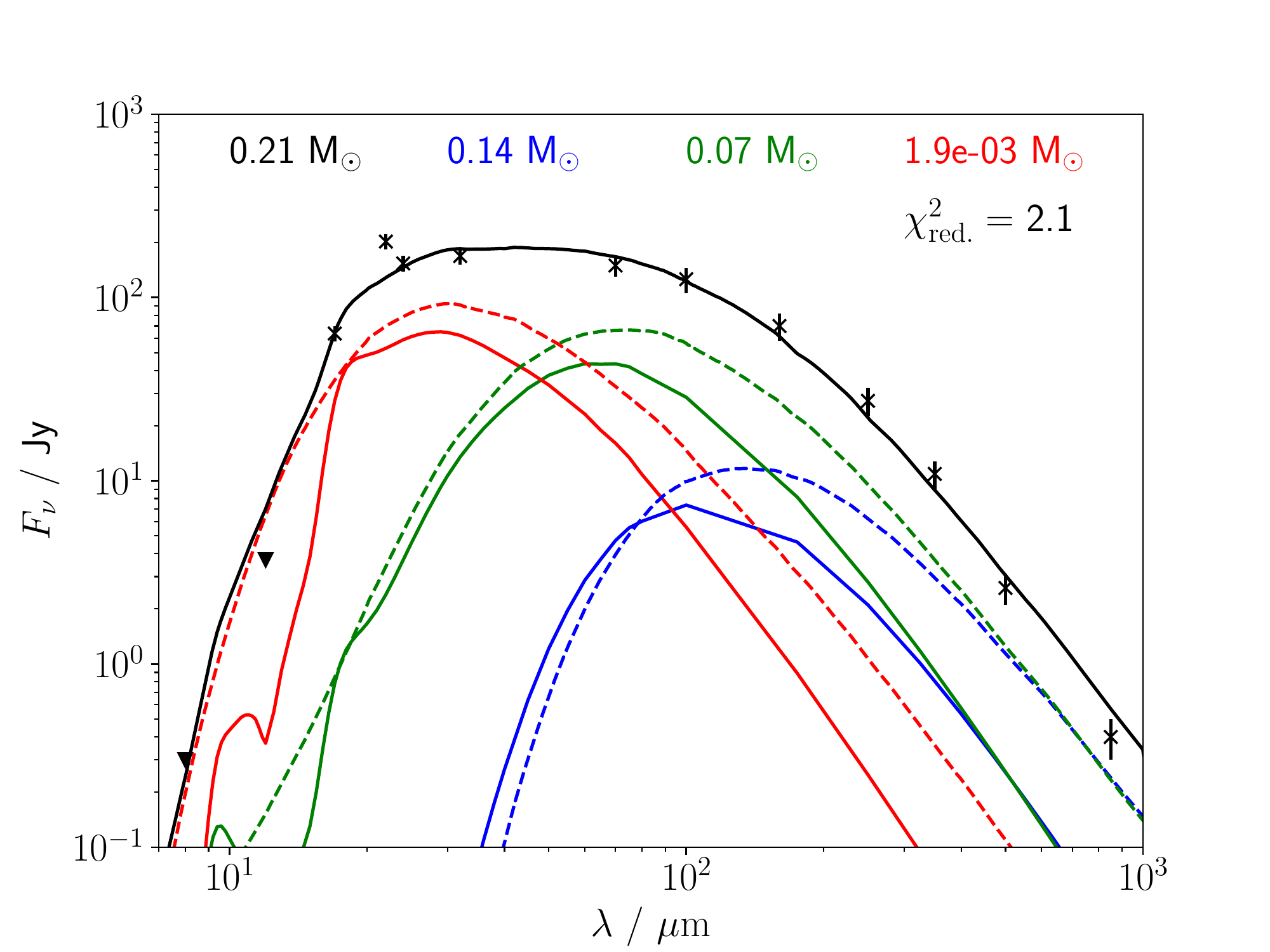}
  \caption{Cas A dust SED (black crosses) and best-fit model SEDs for preshock $0.1 \um$ (blue), clumped $5 \nm$ (green), and diffuse $0.1 \um$ (red) grains, with the total model SED shown in black. The left panel uses Mg$_{2.4}$SiO$_{4.4}$ optical properties (model H), the right panel a $50:50$ mixture of MgSiO$_3$ (solid lines) and amorphous carbon (dashed lines) grains (model I).}
  \label{fig:diffsil}
\end{figure*}

\subsection{Dust masses and grain sizes}

\begin{table*}
  \centering
  \caption{Median dust masses for each component and the best-fit $\chsq$ values for different combinations of grain size, {listed as $a_{\rm preshock}$/$a_{\rm clumped}$/$a_{\rm diffuse}$}. The uncertainties give the $16$th and $84$th percentiles from the MCMC.}
  \begin{tabular}{ccccccc}
    \hline
    & & \multicolumn{4}{c}{Dust mass /$\msun$} & \\
    Model & Grain size /$\um$ & Preshock & Clumped & Diffuse & Total & $\chsq$ \\
    \hline
    A & $1.0$/$1.0$/$1.0$ & $0.10^{+0.72}_{-0.08}$ & $1.00^{+0.14}_{-0.61}$ & $2.4^{+0.1}_{-0.1} \times 10^{-2}$ & $1.17^{+0.10}_{-0.09}$ & $11.3$ \\
    B & $0.01$/$0.01$/$0.01$ & $0.57^{+0.04}_{-0.56}$ & $0.00^{+0.21}_{-0.00}$ & $4.6^{+0.2}_{-0.2} \times 10^{-4}$ & $0.57^{+0.03}_{-0.32}$ & $7.8$ \\
    C & $0.1$/$0.1$/$0.1$ & $0.00^{+0.26}_{-0.00}$ & $0.70^{+0.05}_{-0.22}$ & $2.5^{+0.1}_{-0.1} \times 10^{-3}$ & $0.73^{+0.06}_{-0.04}$ & $1.4$ \\
    D & $0.1$/$0.01$/$0.1$ & $0.65^{+0.11}_{-0.09}$ & $0.069^{+0.023}_{-0.027}$ & $2.5^{+0.1}_{-0.1} \times 10^{-3}$ & $0.72^{+0.08}_{-0.08}$ & $1.2$ \\
    E & $0.1$/$0.005$/$0.1$ & $0.70^{+0.11}_{-0.09}$ & $0.046^{+0.016}_{-0.024}$ & $2.4^{+0.1}_{-0.1} \times 10^{-3}$ & $0.74^{+0.08}_{-0.07}$ & $1.2$ \\
    F & $1.0$/$0.01$/$0.1$ & $0.91^{+0.15}_{-0.12}$ & $0.13^{+0.01}_{-0.02}$ & $2.4^{+0.1}_{-0.1} \times 10^{-3}$ & $1.05^{+0.14}_{-0.11}$ & $1.4$ \\
    G & $1.0$/$0.005$/$0.1$ & $1.04^{+0.17}_{-0.12}$ & $0.094^{+0.011}_{-0.016}$ & $2.3^{+0.1}_{-0.1} \times 10^{-3}$ & $1.13^{+0.16}_{-0.11}$ & $1.6$ \\
    H* & $0.1$/$0.005$/$0.1$ & $0.74^{+0.12}_{-0.09}$ & $0.051^{+0.016}_{-0.023}$ & $2.3^{+0.1}_{-0.1} \times 10^{-3}$ & $0.79^{+0.09}_{-0.08}$ & $0.7$ \\
    I$^\dagger$ & $0.1$/$0.005$/$0.1$ & $0.13^{+0.03}_{-0.03}$ & $0.069^{+0.009}_{-0.010}$ & $1.9^{+0.1}_{-0.1} \times 10^{-3}$ & $0.20^{+0.03}_{-0.03}$ & $2.1$ \\
    \hline
    \multicolumn{7}{c}{*With Mg$_{2.4}$SiO$_{4.4}$ optical properties. $^\dagger$With a $50:50$ MgSiO$_3$:amorphous carbon mixture.}
  \end{tabular}
  \label{tab:dustmass}
\end{table*}

{Figure \ref{fig:mrn} shows the best-fit model assuming a \citet{mathis1977} size distribution for each component, as in \citet{priestley2019}. The model fails to reproduce the observed mid-IR fluxes, as the high-temperature small grains which make up most of the distribution emit too strongly at $\sim 10 \um$, exceeding the observational upper limits\footnote{The $8/12 \um$ fluxes were not included in the fitting procedure in \citet{priestley2019}. It can be seen in that paper that the predicted model fluxes do, in fact, exceed these values in all cases.}. It is thus necessary to consider the grain size in each component individually.}

{Figure \ref{fig:diffsizes} shows best-fit models for grain sizes of either $1 \um$ (model A) or $10 \nm$ (model B) in all three components; median dust masses with uncertainties are given in Table \ref{tab:dustmass}. Both models fail to fit the mid-IR SED, with grain temperatures being too low or too high for models A and B respectively. The observed SED shape in the mid-IR requires grains with a size $\sim 0.1 \um$ in the diffuse component.} Figure \ref{fig:all01} shows the best-fit model for $0.1 \um$ grains in all three components {(model C)}, with $0.72 \msun$ of post-shock clumped dust, $2.5 \times 10^{-3} \msun$ in the diffuse phase, and a negligible quantity of unshocked dust. {While model C is a good fit to the data, it is physically problematic,} as it would imply formation, rather than destruction, of dust in the reverse shock\footnote{While there are some observational \citep{matsuura2019} and theoretical \citep{kirchschlager2020} justifications for post-shock dust reformation, these are to a far lesser extent than implied by the best-fit model.}. The majority of the dust is observed to lie within the reverse-shock radius \citep{delooze2017}, and extinction-based measurements also require a significant mass of unshocked ejecta dust \citep{bevan2017,niculescu2021}. A clumped dust grain size of $0.1 \um$ is also in conflict with the post-shock SED, shown in Figure \ref{fig:postshock}, which clearly requires smaller, higher-temperature grains in this phase.

{To maintain physical consistency, we require that both the DTG ratios and the grain sizes in the two post-shock components are smaller than those in the preshock component, thereby assuming that accretion onto grains or grain coagulation cannot exceed grain destruction in the reverse shock.} Figure \ref{fig:clumpsmall} shows the {results for models with $0.1 \um$ grains in both the preshock and diffuse components, and either $10 \nm$ (model D) or $5 \nm$ (model E) clumped grain sizes.} {Models D and E both} fit the observed $70 \um$ flux better than the all-$0.1 \um$ {model C}, requiring $0.63-0.68 \msun$ of preshock dust, $0.05-0.07 \msun$ of dust in the post-shock clumps, and $2.4 \times 10^{-3} \msun$ of dust in the diffuse material. This is in much better agreement with the inferred {spatial} distribution of the dust mass in the SNR. The smaller grain size in the post-shock clumps is a natural result of shattering via grain-grain collisions \citep{kirchschlager2019}. The grain size in the diffuse phase, {which} is strongly constrained to be $\sim 0.1 \um$ (Figure \ref{fig:diffsizes}), could be due to smaller grains being rapidly destroyed via sputtering in the high-temperature gas, or their being more strongly-coupled to the gas and thus unable to escape from the clumps.

{While we argue that the preshock grain size cannot be smaller than the $\sim 0.1 \um$ in the diffuse component, it could be larger. This would be in better agreement with observations of other SNRs, which are often found to require micron-sized grains \citep[e.g.][]{priestley2020}. Figure \ref{fig:preshockbig} shows best-fit models for a $1 \um$ preshock grain size, with either a $10 \nm$ (model F) or $5 \nm$ (model G) clumped grain size. These models require larger dust masses in both the preshock and clumped components, with a slightly worse $\chsq$ compared to models D and E with $0.1 \um$ preshock grains. \citet{niculescu2021} find that micron-sized grains produce unphysically-large dust masses if responsible for the measured optical extinction in Cas A, and it is unclear whether shattering is efficient enough to completely reprocess a population of $1 \um$ grains into the $0.1 \um$ required in the diffuse component\footnote{Grains this large are almost completely unaffected by sputtering \citep{nozawa2007}.}. We therefore consider models D and E as more plausible, although we cannot rule out the larger grain sizes.}

{We note that no model correctly reproduces the SED peak at $21 \um$. This is an issue of dust composition {in the diffuse component}, which we discuss briefly in Appendix \ref{sec:21um}.} {As this component makes up a negligible fraction of the total dust mass, its importance for the derived destruction efficiency is limited. The assumed silicate composition in the preshock and clumped components also has little effect on our results. Using optical properties from the Mg$_{2.4}$SiO$_{4.4}$ sample in \citet{jaeger2003} (model H), rather than the MgSiO$_3$ grains from \citet{dorschner1995}, and the same grain sizes as model E, we find statistically-indistinguishable dust masses, shown in Figure \ref{fig:diffsil}.}

{If we assume a $50:50$ mixture by mass of MgSiO$_3$ silicate and carbon grains (using ACAR amorphous carbon properties from \citet{zubko1996}, and a bulk density of $1.6 \gcc$; model I), we find a total dust mass a factor of $\sim 4$ lower than for pure-silicate models, and lower than the values reported by \citet{delooze2017} and \citet{bevan2017} for the same mixture by a similar factor. This is due to most of the far-IR flux coming from post-shock carbon grains in our best-fit model, which are warmer (and thus more emissive) than the preshock silicates which provide nearly all the far-IR flux in the other models (a similar effect occurred in the carbon grain models from \citealt{priestley2019}). Model I is a worse fit to the data than our preferred silicate-only models, particularly in the mid-IR, where it exceeds both the $8$ and $12 \um$ upper limits on the flux. Cas A is also an oxygen-rich SNR \citep{docenko2010}, which suggests that silicate grains should be predominant, so we consider model I to be less plausible than the silicate-only models. In any case, the final dust yield for model I obtained below is comparable to those from the silicate-only models (Table \ref{tab:fdest}).}

\subsection{Destruction efficiencies}

\begin{table*}
  \centering
  \caption{Median DTG ratios for the three ejecta components, and the inferred destruction efficiencies in the two post-shock components separately and in combination. The uncertainties give the $16$th and $84$th percentiles from the MCMC. {See Table \ref{tab:dustmass} for model details.}}
  \begin{tabular}{ccccccccc}
    \hline
    & \multicolumn{3}{c}{DTG ratio} & & \multicolumn{3}{c}{$\fdest$} \\
    Model & Preshock & Clumped & Diffuse & & Clumped & Diffuse & Total \\
    \hline
    D & $1.22^{+0.20}_{-0.18}$ & $0.12^{+0.04}_{-0.05}$ & $1.5^{+0.1}_{-0.1} \times 10^{-3}$ & & $0.81^{+0.08}_{-0.07}$ & $0.9973^{+0.0002}_{-0.0002}$ & $0.94^{+0.02}_{-0.02}$ \\
    E & $1.31^{+0.20}_{-0.16}$ & $0.08^{+0.03}_{-0.04}$ & $1.4^{+0.1}_{-0.1} \times 10^{-3}$ & & $0.87^{+0.07}_{-0.05}$ & $0.9975^{+0.0001}_{-0.0002}$ & $0.96^{+0.02}_{-0.01}$ \\
    F & $1.72^{+0.28}_{-0.23}$ & $0.23^{+0.03}_{-0.03}$ & $1.4^{+0.1}_{-0.1} \times 10^{-3}$ & & $0.71^{+0.04}_{-0.04}$ & $0.9977^{+0.0001}_{-0.0001}$ & $0.91^{+0.01}_{-0.01}$ \\
    G & $1.96^{+0.23}_{-0.31}$ & $0.16^{+0.02}_{-0.03}$ & $1.4^{+0.1}_{-0.1} \times 10^{-3}$ & & $0.79^{+0.04}_{-0.03}$ & $0.9979^{+0.0001}_{-0.0001}$ & $0.94^{+0.01}_{-0.01}$ \\
    H & $1.40^{+0.17}_{-0.22}$ & $0.08^{+0.04}_{-0.03}$ & $1.4^{+0.1}_{-0.1} \times 10^{-3}$ & & $0.86^{+0.05}_{-0.06}$ & $0.9976^{+0.0002}_{-0.0001}$ & $0.96^{+0.01}_{-0.02}$ \\
    I & $0.25^{+0.07}_{-0.06}$ & $0.12^{+0.02}_{-0.02}$ & $1.1^{+0.1}_{-0.1} \times 10^{-3}$ & & $0.48^{+0.14}_{-0.20}$ & $0.9945^{+0.0009}_{-0.0014}$ & $0.85^{+0.04}_{-0.06}$ \\
    \hline
  \end{tabular}
  \label{tab:fdest}
\end{table*}

{Combining the dust masses in Table \ref{tab:dustmass} {of models D-I, which are both good fits to the data and physically plausible,} with the corresponding gas masses from Table \ref{tab:gasprop}, we calculate the DTG ratio for each ejecta component, listed in Table \ref{tab:fdest}. We find preshock DTG ratios greater than unity, clumped DTG ratios of $\sim 0.1$ and diffuse DTG ratios $\sim 10^{-3}$. If the clumped and diffuse ejecta initially had the same DTG ratios as the preshock component, this suggests significant and almost total dust destruction in the other two components respectively.}

{The destruction efficiency, $\fdest = 1 - M_f/M_i$ where the subscripts refer to the initial and final dust masses, can be expressed in terms of the DTG ratios\footnote{A derivation is given in Appendix \ref{sec:fdest}.} as
  \begin{equation}
    \fdest = 1 - \frac{{\rm DTG}_f}{{\rm DTG}_i} \, \frac{1 + {\rm DTG}_i}{1 + {\rm DTG}_f}.
  \label{eq:fdest}
  \end{equation}
  Assuming the initial DTG ratio is identical to the preshock component, {for our preferred models with a $0.1 \um$ preshock grain size (D, E and H)} this suggests $\fdest = 74-94 \%$ in the clumped ejecta and $> 99 \%$ in the diffuse component. The inferred total destruction efficiency, accounting for the relative mass in each post-shock component, is $92-98 \%$. Applying the $\fdest$ value for the clumped ejecta to the current (presumably clumped) unshocked dust mass suggests that $0.03-0.21 \msun$ should survive the reverse shock. If no further destruction occurs, then combined with the current post-shock dust mass this gives a total dust yield ($M_{\rm diffuse}+M_{\rm clump} + (1-\fdest)M_{\rm preshock}$) for Cas A of $0.05-0.30 \msun$. If we assume that the diffuse ejecta represents material stripped from clumps during the passage of the reverse shock, then the appropriate value is the total $\fdest$, and the resulting dust yield is $0.03-0.15 \msun$.}

{If the preshock grain size is $1 \um$ (models F and G), then $\fdest$ is lower and both pre- and post-shock dust masses are higher, leading to dust yields approaching $0.5 \msun$. For model I, with a $50:50$ carbon-to-silicate grain ratio, the minimum dust yield is still $\sim 0.12 \msun$, despite the total (current) dust mass being significantly lower than the all-silicate models - the proportion of shocked to unshocked dust mass is much larger than in the silicate models, leading to a low inferred $\fdest$ and a higher relative final yield.}

\section{Discussion}

\subsection{Comparison with theory}

{Our values of $\fdest$ for the clumped ejecta {in models D and E} are somewhat lower than those from {dust destruction} models including grain-grain collisions \citep{kirchschlager2019}, although there is overlap between the ranges. We note that the model clumps in \citet{kirchschlager2019} are mostly dispersed into what would be termed `diffuse' ejecta by our definition, and the total $\fdest$ values accounting for this component are in closer agreement with the theoretical prediction. While \citet{slavin2020} find $\fdest \sim 80-90 \%$ for silicate grains in clumps, close to that observed, this includes significant additional destruction in the ISM after grains have escaped the SNR. The immediate post-shock value of $\fdest$ from \citet{slavin2020} appears to be substantially {lower} than those in Table \ref{tab:fdest}, likely due to their model not accounting for grain-grain collisions \citep{kirchschlager2019}.} {We note that \citet{delooze2017} estimated $\fdest \sim 70 \%$ from the spatial distribution of the dust mass and the radius of the reverse shock, consistent with our models D and E and in good agreement with models F and G.}

\subsection{Implications}

{Although our estimated dust yield for Cas A has a fairly large uncertainty, the $\sim 0.1 \msun$ per CCSNe value required to explain high-redshift observations \citep{dwek2007} is well within reach. We note that Cas A is an exceptionally strongly-interacting SNR, and thus presumably has a much higher $\fdest$ and lower dust yield than less extreme objects. This suggests that CCSNe may be major dust producers even up to the present day. \citet{delooze2020} find that this is the case for $\fdest \lesssim 50\%$, not far from our lower limit, which is again for a very strong reverse shock. \citet{galliano2021} find a maximum possible dust yield of $0.03 \msun$ per CCSNe; even our lower limit exceeds this, and for our best-fit models, a larger dust mass has already survived passing through the reverse shock. The \citet{galliano2021} CCSNe yields are primarily constrained by observations of the dust mass in low-metallicity galaxies. If our results are correct, and applicable beyond Cas A, this suggests that {extending dust evolution models towards low-metallicity environments requires a better understanding of how SN dust yields, gaseous inflows/outflows \citep{nanni2020}, and ISM dust destruction efficiencies \citep{priestley2021b} vary with metallicity, rather than assuming that these remain fixed.}

\subsection{High dust-to-gas ratios}

\citet{laming2020} suggest that, due to the preshock gas mass of $\sim 0.6 \msun$, it is implausible that there is also $0.6 \msun$ of dust in the unshocked ejecta, as this implies a DTG ratio of around unity. We disagree with this argument. The ejecta of Cas A is almost entirely made up of elements which form dust \citep{docenko2010}, and such high condensation efficiencies are also seen in other objects; the Crab Nebula contains $\sim 0.2 \msun$ of metals \citep{owen2015} and $0.02-0.08 \msun$ of dust \citep{delooze2019,priestley2020}, while born-again planetary nebulae, which also feature dust formation in explosively-ejected, heavily fusion-processed ejecta, can reach DTG ratios of $\sim 1$ \citep[e.g.][]{toala2021}. As mentioned previously, two independent studies based on optical line extinction \citep{bevan2017,niculescu2021} have found Cas A dust masses even larger than the $\sim 0.6 \msun$ in this and previous IR-based works, so we consider the \citet{laming2020} value for the preshock gas mass as {possible} evidence of highly efficient dust formation in Cas A, rather than as indicating an error in the determination of the dust mass.

\subsection{Caveats}

{The distribution of dust mass between the various ejecta components appears to be fairly robust, as our results here do not greatly differ from those in \citet{priestley2019} despite the different methodologies. The gas masses represent a potentially much larger source of uncertainty not captured by the errors in Table \ref{tab:fdest}. More recent estimates of the mass of X-ray emitting material are comparable with the value from \citet{willingale2003} we use \citep[e.g.][]{hwang2012}, but the unshocked gas masses from \citet{laming2020} and this work, despite agreeing within the errors, are based on conflicting assumptions. \citet{laming2020} assume the unshocked ejecta has density $\sim 10 \pcc$ and temperature $\sim 8000 \kel$, whereas in Equation \ref{eq:mass} we used $T = 100 \kel$ as found by \citet{raymond2018}, and the density of the post-shock clumps suggests a preshock density of $\sim 100 \pcc$ \citep{docenko2010}. In \citet{priestley2019} we also found an average ionisation state of $Z=1$ for the clumped gas, lower than the $Z=3$ assumed for the unshocked ejecta in Equation \ref{eq:mass}. The lower value gives a preshock gas mass of $2.5 \msun$, which then results in $\fdest \sim 50 \%$ {for models D/E} and a correspondingly enhanced dust yield of $\sim 0.5 \msun$. Observations of atomic line emission by the {\it James Webb Space Telescope}, combined with a model accounting for all ejecta components, would be extremely useful in resolving this uncertainty.}

\section{Conclusions}

We have reanalysed the IR SED of Cas A with a substantially improved dust emission model, finding {dust masses of $0.6-0.8 \msun$ in the unshocked material, $0.02-0.09 \msun$ in the post-shock clumps, and $2.3-2.5 \times 10^{-3} \msun$ in the X-ray emitting diffuse gas {for our preferred preshock grain size of $0.1 \um$}. Combined with updated gas mass estimates, these give DTG ratios of $1.0-1.5$ in the unshocked gas, $0.04-0.16$ in the shocked clumps, and $0.001$ in the X-ray emitting gas. The implied dust destruction efficiency of the reverse shock is $74-94 \%$ in the clumped material and $92-98 \%$ overall, with a final dust yield for Cas A of $0.05-0.30 \msun$ that is sufficient to explain the observed dust masses in high-redshift galaxies.} {For a preshock grain size of $1 \um$, the final dust yield may exceed $0.5 \msun$.} As Cas A is one of the most strongly-interacting SNRs known, the large dust yield {even for this object} suggests that CCSNe in general are efficient dust producers, and contribute significantly to the overall cosmic dust budget.

\section*{Acknowledgements}

{We are grateful to Martin Laming for a useful discussion about the gas-phase ejecta properties.} FDP is funded by the Science and Technology Facilities Council. IDL acknowledges support from European Research Council (ERC) starting grant 851622 DustOrigin. MJB acknowledges support from the ERC grant SNDUST ERC-2015-AdG-694520.

\section*{Data Availability}

The data underlying this article will be made available upon request.

%%%%%%%%%%%%%%%%%%%%%%%%%%%%%%%%%%%%%%%%%%%%%%%%%%



\bibliographystyle{mnras}
\bibliography{revshock}

\begin{thebibliography}{}
\makeatletter
\relax
\def\mn@urlcharsother{\let\do\@makeother \do\$\do\&\do\#\do\^\do\_\do\%\do\~}
\def\mn@doi{\begingroup\mn@urlcharsother \@ifnextchar [ {\mn@doi@}
  {\mn@doi@[]}}
\def\mn@doi@[#1]#2{\def\@tempa{#1}\ifx\@tempa\@empty \href
  {http://dx.doi.org/#2} {doi:#2}\else \href {http://dx.doi.org/#2} {#1}\fi
  \endgroup}
\def\mn@eprint#1#2{\mn@eprint@#1:#2::\@nil}
\def\mn@eprint@arXiv#1{\href {http://arxiv.org/abs/#1} {{\tt arXiv:#1}}}
\def\mn@eprint@dblp#1{\href {http://dblp.uni-trier.de/rec/bibtex/#1.xml}
  {dblp:#1}}
\def\mn@eprint@#1:#2:#3:#4\@nil{\def\@tempa {#1}\def\@tempb {#2}\def\@tempc
  {#3}\ifx \@tempc \@empty \let \@tempc \@tempb \let \@tempb \@tempa \fi \ifx
  \@tempb \@empty \def\@tempb {arXiv}\fi \@ifundefined
  {mn@eprint@\@tempb}{\@tempb:\@tempc}{\expandafter \expandafter \csname
  mn@eprint@\@tempb\endcsname \expandafter{\@tempc}}}

\bibitem[\protect\citeauthoryear{{Allen}, {Groves}, {Dopita}, {Sutherland}  \&
  {Kewley}}{{Allen} et~al.}{2008}]{allen2008}
{Allen} M.~G.,  {Groves} B.~A.,  {Dopita} M.~A.,  {Sutherland} R.~S.,
  {Kewley} L.~J.,  2008, \mn@doi [\apjs] {10.1086/589652}, \href
  {https://ui.adsabs.harvard.edu/abs/2008ApJS..178...20A} {178, 20}

\bibitem[\protect\citeauthoryear{{Arias} et~al.,}{{Arias}
  et~al.}{2018}]{arias2018}
{Arias} M.,  et~al., 2018, \mn@doi [\aap] {10.1051/0004-6361/201732411}, \href
  {http://adsabs.harvard.edu/abs/2018A%26A...612A.110A} {612, A110}

\bibitem[\protect\citeauthoryear{{Barlow} et~al.,}{{Barlow}
  et~al.}{2010}]{barlow2010}
{Barlow} M.~J.,  et~al., 2010, \mn@doi [\aap] {10.1051/0004-6361/201014585},
  \href {http://adsabs.harvard.edu/abs/2010A%26A...518L.138B} {518, L138}

\bibitem[\protect\citeauthoryear{{Bertoldi}, {Carilli}, {Cox}, {Fan},
  {Strauss}, {Beelen}, {Omont}  \& {Zylka}}{{Bertoldi}
  et~al.}{2003}]{bertoldi2003}
{Bertoldi} F.,  {Carilli} C.~L.,  {Cox} P.,  {Fan} X.,  {Strauss} M.~A.,
  {Beelen} A.,  {Omont} A.,   {Zylka} R.,  2003, \mn@doi [\aap]
  {10.1051/0004-6361:20030710}, \href
  {http://adsabs.harvard.edu/abs/2003A%26A...406L..55B} {406, L55}

\bibitem[\protect\citeauthoryear{{Bevan} \& {Barlow}}{{Bevan} \&
  {Barlow}}{2016}]{bevan2016}
{Bevan} A.,  {Barlow} M.~J.,  2016, \mn@doi [\mnras] {10.1093/mnras/stv2651},
  \href {http://adsabs.harvard.edu/abs/2016MNRAS.456.1269B} {456, 1269}

\bibitem[\protect\citeauthoryear{{Bevan}, {Barlow}  \& {Milisavljevic}}{{Bevan}
  et~al.}{2017}]{bevan2017}
{Bevan} A.,  {Barlow} M.~J.,   {Milisavljevic} D.,  2017, \mn@doi [\mnras]
  {10.1093/mnras/stw2985}, \href
  {http://adsabs.harvard.edu/abs/2017MNRAS.465.4044B} {465, 4044}

\bibitem[\protect\citeauthoryear{{Bevan} et~al.,}{{Bevan}
  et~al.}{2019}]{bevan2019}
{Bevan} A.,  et~al., 2019, \mn@doi [\mnras] {10.1093/mnras/stz679}, \href
  {https://ui.adsabs.harvard.edu/abs/2019MNRAS.485.5192B} {485, 5192}

\bibitem[\protect\citeauthoryear{{Bevan} et~al.,}{{Bevan}
  et~al.}{2020}]{bevan2020}
{Bevan} A.~M.,  et~al., 2020, \mn@doi [\apj] {10.3847/1538-4357/ab86a2}, \href
  {https://ui.adsabs.harvard.edu/abs/2020ApJ...894..111B} {894, 111}

\bibitem[\protect\citeauthoryear{{Chawner} et~al.,}{{Chawner}
  et~al.}{2019}]{chawner2019}
{Chawner} H.,  et~al., 2019, \mn@doi [\mnras] {10.1093/mnras/sty2942}, \href
  {http://adsabs.harvard.edu/abs/2019MNRAS.483...70C} {483, 70}

\bibitem[\protect\citeauthoryear{{Chawner} et~al.,}{{Chawner}
  et~al.}{2020}]{chawner2020}
{Chawner} H.,  et~al., 2020, \mn@doi [\mnras] {10.1093/mnras/staa221}, \href
  {https://ui.adsabs.harvard.edu/abs/2020MNRAS.493.2706C} {493, 2706}

\bibitem[\protect\citeauthoryear{{De Looze}, {Barlow}, {Swinyard}, {Rho},
  {Gomez}, {Matsuura}  \& {Wesson}}{{De Looze} et~al.}{2017}]{delooze2017}
{De Looze} I.,  {Barlow} M.~J.,  {Swinyard} B.~M.,  {Rho} J.,  {Gomez} H.~L.,
  {Matsuura} M.,   {Wesson} R.,  2017, \mn@doi [\mnras]
  {10.1093/mnras/stw2837}, \href
  {http://adsabs.harvard.edu/abs/2017MNRAS.465.3309D} {465, 3309}

\bibitem[\protect\citeauthoryear{{De Looze} et~al.,}{{De Looze}
  et~al.}{2019}]{delooze2019}
{De Looze} I.,  et~al., 2019, \mn@doi [\mnras] {10.1093/mnras/stz1533}, \href
  {https://ui.adsabs.harvard.edu/abs/2019MNRAS.488..164D} {488, 164}

\bibitem[\protect\citeauthoryear{{De Looze} et~al.,}{{De Looze}
  et~al.}{2020}]{delooze2020}
{De Looze} I.,  et~al., 2020, \mn@doi [\mnras] {10.1093/mnras/staa1496}, \href
  {https://ui.adsabs.harvard.edu/abs/2020MNRAS.496.3668D} {496, 3668}

\bibitem[\protect\citeauthoryear{{DeLaney}, {Kassim}, {Rudnick}  \&
  {Perley}}{{DeLaney} et~al.}{2014}]{delaney2014}
{DeLaney} T.,  {Kassim} N.~E.,  {Rudnick} L.,   {Perley} R.~A.,  2014, \mn@doi
  [\apj] {10.1088/0004-637X/785/1/7}, \href
  {http://adsabs.harvard.edu/abs/2014ApJ...785....7D} {785, 7}

\bibitem[\protect\citeauthoryear{{Docenko} \& {Sunyaev}}{{Docenko} \&
  {Sunyaev}}{2010}]{docenko2010}
{Docenko} D.,  {Sunyaev} R.~A.,  2010, \mn@doi [\aap]
  {10.1051/0004-6361/200810366}, \href
  {http://adsabs.harvard.edu/abs/2010A%26A...509A..59D} {509, A59}

\bibitem[\protect\citeauthoryear{{Dorschner}, {Begemann}, {Henning}, {Jaeger}
  \& {Mutschke}}{{Dorschner} et~al.}{1995}]{dorschner1995}
{Dorschner} J.,  {Begemann} B.,  {Henning} T.,  {Jaeger} C.,   {Mutschke} H.,
  1995, \aap, \href {http://adsabs.harvard.edu/abs/1995A%26A...300..503D} {300,
  503}

\bibitem[\protect\citeauthoryear{{Dwek}, {Galliano}  \& {Jones}}{{Dwek}
  et~al.}{2007}]{dwek2007}
{Dwek} E.,  {Galliano} F.,   {Jones} A.~P.,  2007, \mn@doi [\apj]
  {10.1086/518430}, \href {http://adsabs.harvard.edu/abs/2007ApJ...662..927D}
  {662, 927}

\bibitem[\protect\citeauthoryear{{Foreman-Mackey}, {Hogg}, {Lang}  \&
  {Goodman}}{{Foreman-Mackey} et~al.}{2013}]{foreman2013}
{Foreman-Mackey} D.,  {Hogg} D.~W.,  {Lang} D.,   {Goodman} J.,  2013, \mn@doi
  [\pasp] {10.1086/670067}, \href
  {http://adsabs.harvard.edu/abs/2013PASP..125..306F} {125, 306}

\bibitem[\protect\citeauthoryear{{Gall} et~al.,}{{Gall}
  et~al.}{2014}]{gall2014}
{Gall} C.,  et~al., 2014, \mn@doi [\nat] {10.1038/nature13558}, \href
  {http://adsabs.harvard.edu/abs/2014Natur.511..326G} {511, 326}

\bibitem[\protect\citeauthoryear{{Galliano} et~al.,}{{Galliano}
  et~al.}{2021}]{galliano2021}
{Galliano} F.,  et~al., 2021, \mn@doi [\aap] {10.1051/0004-6361/202039701},
  \href {https://ui.adsabs.harvard.edu/abs/2021A&A...649A..18G} {649, A18}

\bibitem[\protect\citeauthoryear{{Hwang} \& {Laming}}{{Hwang} \&
  {Laming}}{2012}]{hwang2012}
{Hwang} U.,  {Laming} J.~M.,  2012, \mn@doi [\apj]
  {10.1088/0004-637X/746/2/130}, \href
  {https://ui.adsabs.harvard.edu/abs/2012ApJ...746..130H} {746, 130}

\bibitem[\protect\citeauthoryear{{Isensee}, {Rudnick}, {DeLaney}, {Smith},
  {Rho}, {Reach}, {Kozasa}  \& {Gomez}}{{Isensee} et~al.}{2010}]{isensee2010}
{Isensee} K.,  {Rudnick} L.,  {DeLaney} T.,  {Smith} J.~D.,  {Rho} J.,  {Reach}
  W.~T.,  {Kozasa} T.,   {Gomez} H.,  2010, \mn@doi [\apj]
  {10.1088/0004-637X/725/2/2059}, \href
  {https://ui.adsabs.harvard.edu/abs/2010ApJ...725.2059I} {725, 2059}

\bibitem[\protect\citeauthoryear{{J{\"a}ger}, {Dorschner}, {Mutschke}, {Posch}
  \& {Henning}}{{J{\"a}ger} et~al.}{2003}]{jaeger2003}
{J{\"a}ger} C.,  {Dorschner} J.,  {Mutschke} H.,  {Posch} T.,   {Henning} T.,
  2003, \mn@doi [\aap] {10.1051/0004-6361:20030916}, \href
  {http://adsabs.harvard.edu/abs/2003A%26A...408..193J} {408, 193}

\bibitem[\protect\citeauthoryear{{Kirchschlager}, {Schmidt}, {Barlow},
  {Fogerty}, {Bevan}  \& {Priestley}}{{Kirchschlager}
  et~al.}{2019}]{kirchschlager2019}
{Kirchschlager} F.,  {Schmidt} F.~D.,  {Barlow} M.~J.,  {Fogerty} E.~L.,
  {Bevan} A.,   {Priestley} F.~D.,  2019, \mn@doi [\mnras]
  {10.1093/mnras/stz2399}, \href
  {https://ui.adsabs.harvard.edu/abs/2019MNRAS.489.4465K} {489, 4465}

\bibitem[\protect\citeauthoryear{{Kirchschlager}, {Barlow}  \&
  {Schmidt}}{{Kirchschlager} et~al.}{2020}]{kirchschlager2020}
{Kirchschlager} F.,  {Barlow} M.~J.,   {Schmidt} F.~D.,  2020, \mn@doi [\apj]
  {10.3847/1538-4357/ab7db8}, \href
  {https://ui.adsabs.harvard.edu/abs/2020ApJ...893...70K} {893, 70}

\bibitem[\protect\citeauthoryear{{Koo}, {Kim}, {Lee}, {Raymond}, {Lee}, {Yoon}
  \& {Moon}}{{Koo} et~al.}{2018}]{koo2018}
{Koo} B.-C.,  {Kim} H.-J.,  {Lee} Y.-H.,  {Raymond} J.~C.,  {Lee} J.-J.,
  {Yoon} S.-C.,   {Moon} D.-S.,  2018, \mn@doi [\apj]
  {10.3847/1538-4357/aae20e}, \href
  {https://ui.adsabs.harvard.edu/abs/2018ApJ...866..139K} {866, 139}

\bibitem[\protect\citeauthoryear{{Laming} \& {Temim}}{{Laming} \&
  {Temim}}{2020}]{laming2020}
{Laming} J.~M.,  {Temim} T.,  2020, \mn@doi [\apj] {10.3847/1538-4357/abc1e5},
  \href {https://ui.adsabs.harvard.edu/abs/2020ApJ...904..115L} {904, 115}

\bibitem[\protect\citeauthoryear{{Laor} \& {Draine}}{{Laor} \&
  {Draine}}{1993}]{laor1993}
{Laor} A.,  {Draine} B.~T.,  1993, \mn@doi [\apj] {10.1086/172149}, \href
  {http://adsabs.harvard.edu/abs/1993ApJ...402..441L} {402, 441}

\bibitem[\protect\citeauthoryear{{Mart{\'\i}nez-Gonz{\'a}lez}, {W{\"u}nsch},
  {Silich}, {Tenorio-Tagle}, {Palou{\v{s}}}  \&
  {Ferrara}}{{Mart{\'\i}nez-Gonz{\'a}lez} et~al.}{2019}]{martinezgonzales2019}
{Mart{\'\i}nez-Gonz{\'a}lez} S.,  {W{\"u}nsch} R.,  {Silich} S.,
  {Tenorio-Tagle} G.,  {Palou{\v{s}}} J.,   {Ferrara} A.,  2019, \mn@doi [\apj]
  {10.3847/1538-4357/ab571b}, \href
  {https://ui.adsabs.harvard.edu/abs/2019ApJ...887..198M} {887, 198}

\bibitem[\protect\citeauthoryear{{Mathis}, {Rumpl}  \& {Nordsieck}}{{Mathis}
  et~al.}{1977}]{mathis1977}
{Mathis} J.~S.,  {Rumpl} W.,   {Nordsieck} K.~H.,  1977, \mn@doi [\apj]
  {10.1086/155591}, \href {http://adsabs.harvard.edu/abs/1977ApJ...217..425M}
  {217, 425}

\bibitem[\protect\citeauthoryear{{Matsuura} et~al.,}{{Matsuura}
  et~al.}{2015}]{matsuura2015}
{Matsuura} M.,  et~al., 2015, \mn@doi [\apj] {10.1088/0004-637X/800/1/50},
  \href {http://adsabs.harvard.edu/abs/2015ApJ...800...50M} {800, 50}

\bibitem[\protect\citeauthoryear{{Matsuura} et~al.,}{{Matsuura}
  et~al.}{2019}]{matsuura2019}
{Matsuura} M.,  et~al., 2019, \mn@doi [\mnras] {10.1093/mnras/sty2734}, \href
  {https://ui.adsabs.harvard.edu/abs/2019MNRAS.482.1715M} {482, 1715}

\bibitem[\protect\citeauthoryear{{Milisavljevic} \& {Fesen}}{{Milisavljevic} \&
  {Fesen}}{2015}]{milisavljevic2015}
{Milisavljevic} D.,  {Fesen} R.~A.,  2015, \mn@doi [Science]
  {10.1126/science.1261949}, \href
  {https://ui.adsabs.harvard.edu/abs/2015Sci...347..526M} {347, 526}

\bibitem[\protect\citeauthoryear{{Morgan} \& {Edmunds}}{{Morgan} \&
  {Edmunds}}{2003}]{morgan2003}
{Morgan} H.~L.,  {Edmunds} M.~G.,  2003, \mn@doi [\mnras]
  {10.1046/j.1365-8711.2003.06681.x}, \href
  {http://adsabs.harvard.edu/abs/2003MNRAS.343..427M} {343, 427}

\bibitem[\protect\citeauthoryear{{Nanni}, {Burgarella}, {Theul{\'e}},
  {C{\^o}t{\'e}}  \& {Hirashita}}{{Nanni} et~al.}{2020}]{nanni2020}
{Nanni} A.,  {Burgarella} D.,  {Theul{\'e}} P.,  {C{\^o}t{\'e}} B.,
  {Hirashita} H.,  2020, \mn@doi [\aap] {10.1051/0004-6361/202037833}, \href
  {https://ui.adsabs.harvard.edu/abs/2020A&A...641A.168N} {641, A168}

\bibitem[\protect\citeauthoryear{{Niculescu-Duvaz}, {Barlow}, {Bevan},
  {Milisavljevic}  \& {De Looze}}{{Niculescu-Duvaz}
  et~al.}{2021}]{niculescu2021}
{Niculescu-Duvaz} M.,  {Barlow} M.~J.,  {Bevan} A.,  {Milisavljevic} D.,   {De
  Looze} I.,  2021, \mn@doi [\mnras] {10.1093/mnras/stab932}, \href
  {https://ui.adsabs.harvard.edu/abs/2021MNRAS.504.2133N} {504, 2133}

\bibitem[\protect\citeauthoryear{{Nozawa}, {Kozasa}, {Habe}, {Dwek}, {Umeda},
  {Tominaga}, {Maeda}  \& {Nomoto}}{{Nozawa} et~al.}{2007}]{nozawa2007}
{Nozawa} T.,  {Kozasa} T.,  {Habe} A.,  {Dwek} E.,  {Umeda} H.,  {Tominaga} N.,
   {Maeda} K.,   {Nomoto} K.,  2007, \mn@doi [\apj] {10.1086/520621}, \href
  {http://adsabs.harvard.edu/abs/2007ApJ...666..955N} {666, 955}

\bibitem[\protect\citeauthoryear{{Owen} \& {Barlow}}{{Owen} \&
  {Barlow}}{2015}]{owen2015}
{Owen} P.~J.,  {Barlow} M.~J.,  2015, \mn@doi [\apj]
  {10.1088/0004-637X/801/2/141}, \href
  {http://adsabs.harvard.edu/abs/2015ApJ...801..141O} {801, 141}

\bibitem[\protect\citeauthoryear{{Priddey}, {Isaak}, {McMahon}, {Robson}  \&
  {Pearson}}{{Priddey} et~al.}{2003}]{priddey2003}
{Priddey} R.~S.,  {Isaak} K.~G.,  {McMahon} R.~G.,  {Robson} E.~I.,   {Pearson}
  C.~P.,  2003, \mn@doi [\mnras] {10.1046/j.1365-8711.2003.07076.x}, \href
  {http://adsabs.harvard.edu/abs/2003MNRAS.344L..74P} {344, L74}

\bibitem[\protect\citeauthoryear{{Priestley}, {Barlow}  \& {De
  Looze}}{{Priestley} et~al.}{2019}]{priestley2019}
{Priestley} F.~D.,  {Barlow} M.~J.,   {De Looze} I.,  2019, \mn@doi [\mnras]
  {10.1093/mnras/stz414}, \href
  {http://adsabs.harvard.edu/abs/2019MNRAS.485..440P} {485, 440}

\bibitem[\protect\citeauthoryear{{Priestley}, {Barlow}, {De Looze}  \&
  {Chawner}}{{Priestley} et~al.}{2020}]{priestley2020}
{Priestley} F.~D.,  {Barlow} M.~J.,  {De Looze} I.,   {Chawner} H.,  2020,
  \mn@doi [\mnras] {10.1093/mnras/stz3434}, \href
  {https://ui.adsabs.harvard.edu/abs/2020MNRAS.491.6020P} {491, 6020}

\bibitem[\protect\citeauthoryear{{Priestley}, {De Looze}  \&
  {Barlow}}{{Priestley} et~al.}{2021}]{priestley2021b}
{Priestley} F.~D.,  {De Looze} I.,   {Barlow} M.~J.,  2021, arXiv e-prints,
  \href {https://ui.adsabs.harvard.edu/abs/2021arXiv211006952P} {p.
  arXiv:2110.06952}

\bibitem[\protect\citeauthoryear{{Raymond}, {Koo}, {Lee}, {Milisavljevic},
  {Fesen}  \& {Chilingarian}}{{Raymond} et~al.}{2018}]{raymond2018}
{Raymond} J.~C.,  {Koo} B.-C.,  {Lee} Y.-H.,  {Milisavljevic} D.,  {Fesen}
  R.~A.,   {Chilingarian} I.,  2018, \mn@doi [\apj] {10.3847/1538-4357/aadf93},
  \href {http://adsabs.harvard.edu/abs/2018ApJ...866..128R} {866, 128}

\bibitem[\protect\citeauthoryear{{Reed}, {Hester}, {Fabian}  \&
  {Winkler}}{{Reed} et~al.}{1995}]{reed1995}
{Reed} J.~E.,  {Hester} J.~J.,  {Fabian} A.~C.,   {Winkler} P.~F.,  1995,
  \mn@doi [\apj] {10.1086/175308}, \href
  {http://adsabs.harvard.edu/abs/1995ApJ...440..706R} {440, 706}

\bibitem[\protect\citeauthoryear{{Rho} et~al.,}{{Rho} et~al.}{2008}]{rho2008}
{Rho} J.,  et~al., 2008, \mn@doi [\apj] {10.1086/523835}, \href
  {http://adsabs.harvard.edu/abs/2008ApJ...673..271R} {673, 271}

\bibitem[\protect\citeauthoryear{{Rho} et~al.,}{{Rho} et~al.}{2018}]{rho2018}
{Rho} J.,  et~al., 2018, \mn@doi [\mnras] {10.1093/mnras/sty1713}, \href
  {http://adsabs.harvard.edu/abs/2018MNRAS.479.5101R} {479, 5101}

\bibitem[\protect\citeauthoryear{{Slavin}, {Dwek}, {Mac Low}  \&
  {Hill}}{{Slavin} et~al.}{2020}]{slavin2020}
{Slavin} J.~D.,  {Dwek} E.,  {Mac Low} M.-M.,   {Hill} A.~S.,  2020, \mn@doi
  [\apj] {10.3847/1538-4357/abb5a4}, \href
  {https://ui.adsabs.harvard.edu/abs/2020ApJ...902..135S} {902, 135}

\bibitem[\protect\citeauthoryear{{Smith}, {Rudnick}, {Delaney}, {Rho}, {Gomez},
  {Kozasa}, {Reach}  \& {Isensee}}{{Smith} et~al.}{2009}]{smith2009}
{Smith} J.~D.~T.,  {Rudnick} L.,  {Delaney} T.,  {Rho} J.,  {Gomez} H.,
  {Kozasa} T.,  {Reach} W.,   {Isensee} K.,  2009, \mn@doi [\apj]
  {10.1088/0004-637X/693/1/713}, \href
  {http://adsabs.harvard.edu/abs/2009ApJ...693..713S} {693, 713}

\bibitem[\protect\citeauthoryear{{Toal{\'a}} et~al.,}{{Toal{\'a}}
  et~al.}{2021}]{toala2021}
{Toal{\'a}} J.~A.,  et~al., 2021, \mn@doi [\mnras] {10.1093/mnras/stab593},
  \href {https://ui.adsabs.harvard.edu/abs/2021MNRAS.503.1543T} {503, 1543}

\bibitem[\protect\citeauthoryear{{Wang} \& {Li}}{{Wang} \&
  {Li}}{2016}]{wang2016}
{Wang} W.,  {Li} Z.,  2016, \mn@doi [\apj] {10.3847/0004-637X/825/2/102}, \href
  {http://adsabs.harvard.edu/abs/2016ApJ...825..102W} {825, 102}

\bibitem[\protect\citeauthoryear{{Watson}, {Christensen}, {Knudsen}, {Richard},
  {Gallazzi}  \& {Micha{\l}owski}}{{Watson} et~al.}{2015}]{watson2015}
{Watson} D.,  {Christensen} L.,  {Knudsen} K.~K.,  {Richard} J.,  {Gallazzi}
  A.,   {Micha{\l}owski} M.~J.,  2015, \mn@doi [\nat] {10.1038/nature14164},
  \href {http://adsabs.harvard.edu/abs/2015Natur.519..327W} {519, 327}

\bibitem[\protect\citeauthoryear{{Wesson}, {Barlow}, {Matsuura}  \&
  {Ercolano}}{{Wesson} et~al.}{2015}]{wesson2015}
{Wesson} R.,  {Barlow} M.~J.,  {Matsuura} M.,   {Ercolano} B.,  2015, \mn@doi
  [\mnras] {10.1093/mnras/stu2250}, \href
  {http://adsabs.harvard.edu/abs/2015MNRAS.446.2089W} {446, 2089}

\bibitem[\protect\citeauthoryear{{Willingale}, {Bleeker}, {van der Heyden}  \&
  {Kaastra}}{{Willingale} et~al.}{2003}]{willingale2003}
{Willingale} R.,  {Bleeker} J.~A.~M.,  {van der Heyden} K.~J.,   {Kaastra}
  J.~S.,  2003, \mn@doi [\aap] {10.1051/0004-6361:20021554}, \href
  {http://adsabs.harvard.edu/abs/2003A%26A...398.1021W} {398, 1021}

\bibitem[\protect\citeauthoryear{{Zubko}, {Mennella}, {Colangeli}  \&
  {Bussoletti}}{{Zubko} et~al.}{1996}]{zubko1996}
{Zubko} V.~G.,  {Mennella} V.,  {Colangeli} L.,   {Bussoletti} E.,  1996,
  \mn@doi [\mnras] {10.1093/mnras/282.4.1321}, \href
  {http://adsabs.harvard.edu/abs/1996MNRAS.282.1321Z} {282, 1321}

\bibitem[\protect\citeauthoryear{{de Gasperin} et~al.,}{{de Gasperin}
  et~al.}{2020}]{degasperin2020}
{de Gasperin} F.,  et~al., 2020, \mn@doi [\aap] {10.1051/0004-6361/201936844},
  \href {https://ui.adsabs.harvard.edu/abs/2020A&A...635A.150D} {635, A150}

\makeatother
\end{thebibliography}

%%%%%%%%%%%%%%%%%%%%%%%%%%%%%%%%%%%%%%%%%%%%%%%%%%

%%%%%%%%%%%%%%%%% APPENDICES %%%%%%%%%%%%%%%%%%%%%
\appendix

\section{The $21 \um$ excess}
\label{sec:21um}

\begin{figure}
  \centering
  \includegraphics[width=\columnwidth]{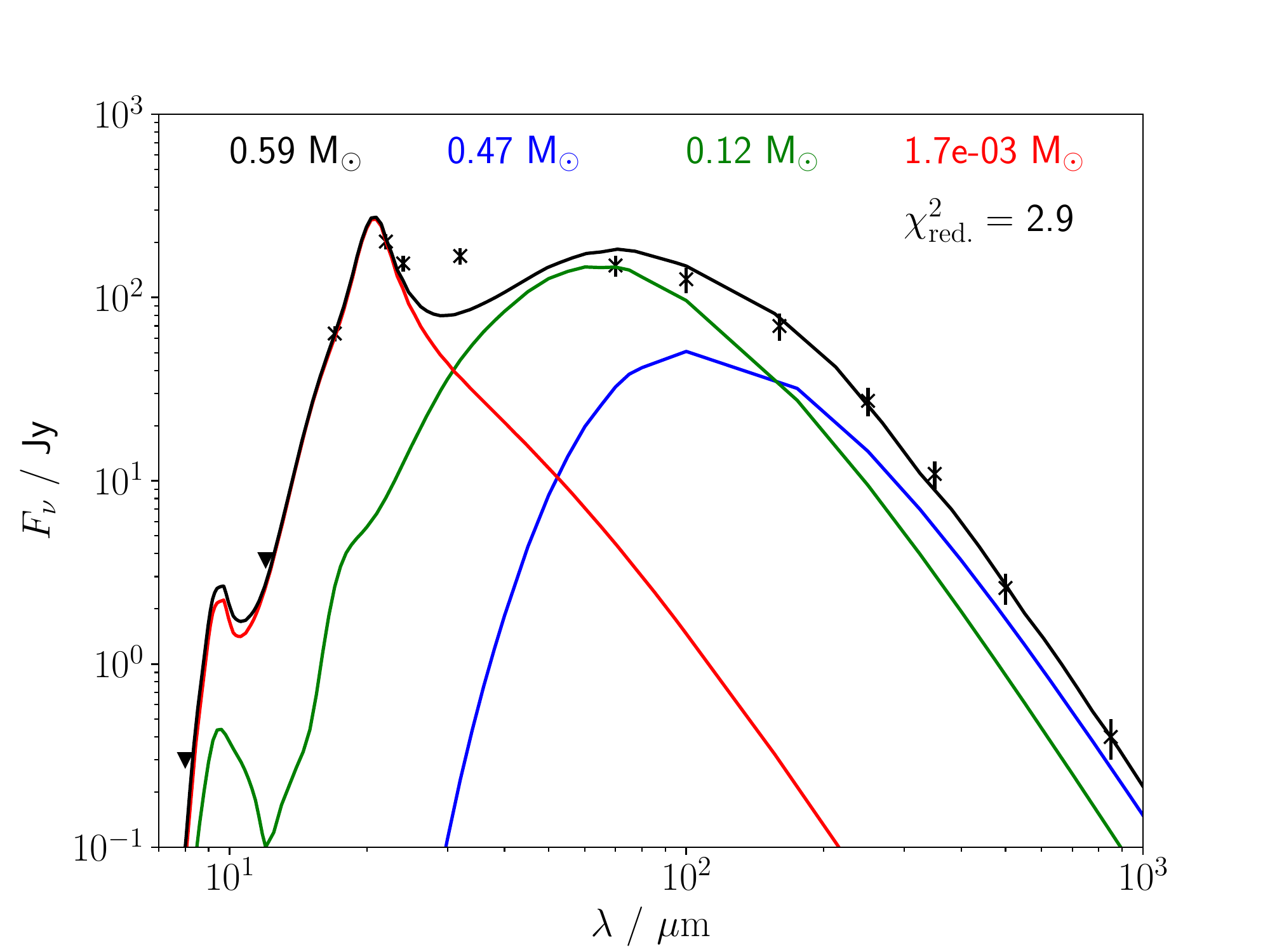}
  \caption{Cas A dust SED (black crosses) and best-fit model SEDs for preshock $0.1 \um$ (blue), clumped $5 \nm$ (green), and diffuse $0.1 \um$ (red) grains, with the total model SED shown in black. The preshock and clumped dust has MgSiO$_3$ optical properties from \citet{dorschner1995}, the diffuse dust Mg$_{0.7}$SiO$_{2.7}$ from \citet{jaeger2003}.}
  \label{fig:21um}
\end{figure}

While grains with MgSiO$_3$ optical properties result in a good fit to nearly all of the IR SED, there is a notable excess at $21 \um$ which is not well-reproduced. This is even more apparent in the \textit{Spitzer} IRS spectrum \citep{rho2008}, and can be fit with a variety of other dust compositions including SiO$_2$ and FeO. A similar excess is also seen in \gfiftyfour{} \citep{rho2018}. The optical properties of Mg$_{0.7}$SiO$_{2.7}$ from \citet{jaeger2003} can reproduce this peak reasonably well, and also do so for \gfiftyfour{} \citep{priestley2020}, but the sharply-declining opacity beyond $21 \um$ is in conflict with the data at these wavelengths (Figure \ref{fig:21um}). Most other proposed carriers of the feature also display this behaviour, suggesting that at least two dust species are present in the diffuse component. This behaviour results in implausibly high dust masses if the same species is used to fit the far-IR SED, which implies that the $21 \um$ carrier is either not present in the colder gas components, or makes up only a small fraction of the total dust mass. We are unable to satisfactorily fit the mid-IR SED using a combination of different silicate grains in the diffuse component; given the uncertainties regarding both dust and gas properties, it is almost certainly possible to achieve this while remaining physically consistent with other observations, but as the dust masses are a small proportion of the total, we do not consider it necessary for the objectives of this paper.

\section{$\fdest$ in terms of DTG ratios}
\label{sec:fdest}

{The dust destruction efficiency, $\fdest = 1 - M_{f,d}/M_{i,d}$, can be expressed in terms of the pre- and post-shock DTG ratios as follows. We multiply the numerator and denominator of the dust mass ratio by $M_{\rm tot} = M_d + M_g$ {($M_{f,d} + M_{f,g} = M_{i,d} + M_{i,g}$, as mass is only transferred between the gas/dust phases, not lost)} to obtain}
\begin{equation}
  \fdest = 1 - \frac{M_{f,d}}{M_{f,d}+M_{f,g}} \, \frac{M_{i,d}+M_{i,g}}{M_{i,d}},
  \label{eq:deriv}
\end{equation}
{and divide {the numerator and denominator of each ratio in Equation \ref{eq:deriv} by the respective value of $M_g$,} to get}
\begin{eqnarray}
  \fdest & = & 1 - \frac{M_{f,d}/M_{f,g}}{1 + M_{f,d}/M_{f,g}} \, \frac{1 + M_{i,d}/M_{i,g}}{M_{i,d}/M_{i,g}} \\
  & = & 1 - \frac{{\rm DTG}_f}{1 + {\rm DTG}_f} \, \frac{1 + {\rm DTG}_i}{{\rm DTG}_i} \\
  & = & 1 - \frac{{\rm DTG}_f}{{\rm DTG}_i} \, \frac{1 + {\rm DTG}_i}{1 + {\rm DTG}_f}
\end{eqnarray}
{as in Equation \ref{eq:fdest}. Note that in \citet{priestley2019} we assumed $\fdest = 1 - {\rm DTG}_f/{\rm DTG_i}$. While this is a reasonable approximation when the DTG ratio is $\ll 1$, as the preshock dust and gas masses in Cas A are comparable this will give erroneous results. For these values, a significant fraction of the current post-shock gas mass was initially locked up in dust grains, so the second term, $(1 + {\rm DTG}_i)/(1 + {\rm DTG}_f)$, cannot be neglected.}

%%%%%%%%%%%%%%%%%%%%%%%%%%%%%%%%%%%%%%%%%%%%%%%%%%

% Don't change these lines
\bsp	% typesetting comment
\label{lastpage}
\end{document}